\documentclass[
 reprint,
 amsmath,amssymb,
 aps,
 floatfix,
]{revtex4-2}

\usepackage{caption}
\usepackage{hyperref}
\usepackage{amsmath}
\usepackage{array}
\usepackage{physics}
\usepackage{amssymb}
\usepackage{braket}
\usepackage{graphicx}
\usepackage{tikz}
\usepackage{dcolumn}
\usepackage{bm}
\usepackage[final,protrusion=true,expansion=true]{microtype}

\DeclareCaptionFormat{paragraphindent}{%
  \parbox{\linewidth}{%
    \leftskip=0pt
    \rightskip=0pt
    \parfillskip=0pt plus 1fil
    \setlength{\parindent}{1.5em}%
    \indent#1#2#3\par}%
}
\graphicspath{{images revised/}}

\makeatletter
\def\pre@bibdata{\jobname\bibdata@app}
\makeatother
\microtypesetup{protrusion=true,expansion=true}
\begin{document}

\preprint{APS/123-QED}

\title{Open-System Probe Selection and Optimal Measurements for Dispersive Coupling Metrology}

\author{Achal Vinod}
\author{Da-Wei Luo}
\author{Ting Yu}
\affiliation{Department of Physics, Stevens Institute of Technology, Hoboken, New Jersey, 07030, USA}

\date{\today}

\begin{abstract}
We investigate how photon loss and environmental memory affect probe selection and
measurement sensitivity for estimating the dispersive coupling strength in a
cavity--qubit system. Using density-operator and quantum-state-diffusion approaches,
we examine the quantum Fisher information (QFI) for representative separable and
entangled initial states in Markovian and non-Markovian regimes. The results show that
closed-system generator variance alone does not determine open-system performance: the
survival of parameter-sensitive coherences and the influence of environmental memory
also play important roles. Within the single-photon preparations examined, a separable
probe can outperform the entangled family considered. Comparisons of selected coherent
and squeezed-vacuum preparations at equal mean photon number further illustrate how
photon-number statistics and qubit preparation affect sensitivity. For the parameters
studied, finite environmental memory can enhance and prolong sensitivity and yield
nonzero QFI for a probe whose Markovian QFI vanishes. We also numerically identify a
fixed Hermitian observable whose error-propagation uncertainty remains close to the
quantum Cram\'er--Rao bound over a prescribed coupling interval in the Markovian regime.
These findings clarify the interplay between probe preparation, environmental
dynamics, and measurement choice in dispersive quantum metrology.

\begin{description}
\item[Keywords]
Cavity QED, quantum Fisher information, non-Markovian dynamics, optimal measurement
\end{description}
\end{abstract}

\maketitle
\raggedbottom

\section{Introduction}

Quantum Fisher information (QFI) determines the ultimate local precision bound for
estimating a parameter encoded in a quantum state
\cite{rao1945information,helstrom1967detection,helstrom1969quantum,braunstein1994statistical,paris2009quantum}.
It is therefore a central tool in quantum metrology
\cite{giovannetti2006quantum,giovannetti2011advances,toth2014quantum}. For unitary
Hamiltonian parameter estimation, the optimal closed-system probe has a simple
structure. If the parameter is generated by a Hermitian operator \(G\), the QFI is
proportional to the variance of \(G\), and the maximum variance is obtained from an
equal-weight superposition of the eigenvectors associated with the smallest and largest
eigenvalues \cite{pang2014quantum}. This result provides a useful benchmark, but it
does not by itself determine the optimal probe for realistic cavity-QED metrology.

The central question addressed here is whether this closed-system optimality principle
survives in an open cavity-qubit system. In the presence of environmental coupling,
different components of an otherwise optimal superposition may decay at different
rates, and noise can impose fundamental precision limits that change the probe-state
optimization problem relative to the closed-system setting
\cite{escher2011general,demkowicz2012elusive}. Markovian loss can remove the coherence
that carries information about the parameter, while environmental memory can produce
revivals of coherence and correlations
\cite{lindblad1976generators,breuer2009measure,budini2018maximally,budini2022quantum,breuer2016colloquium}.
Non-Markovian dynamics have also been connected to the flow of QFI
\cite{chin2012quantum,lu2010quantum}, although memory effects do not universally
improve sensitivity and their interpretation depends on how the estimated parameter
enters the dynamical map.
Consequently, the state with the largest initial generator variance need not yield the
largest usable sensitivity after finite-time open-system evolution.

To resolve this distinction, we employ two complementary QFI diagnostics. The
density-operator QFI gives the precision bound associated with the averaged reduced
state. The stochastic QFI (SQFI), evaluated for individual quantum-state-diffusion
trajectories before the noise average, retains the conditional information carried by
each noise realization. This trajectory-resolved description is useful because the
noise record can contain information that is lost when only the unconditional density
operator is retained; as a result, the averaged SQFI may differ slightly from the
density-operator QFI while remaining consistent with the same open-system dynamics.
Continuous monitoring of the environment can provide access to additional
metrological information, as demonstrated in monitored noisy quantum-metrology
protocols~\cite{albarelli2018restoring}. Such operational access requires an
appropriate monitoring scheme; a formal QSD trajectory should not automatically be
identified with an experimentally accessible measurement record.

We apply this framework to estimation of the dispersive coupling strength \(\chi\) in a
cavity-qubit system. Dispersive interactions in cavity and circuit QED provide a
natural platform for quantum non-demolition readout, photon-number-resolved
spectroscopy, and conditional oscillator control
\cite{haroche1989cavity,blais2004cavity,wallraff2004strong,kono2018quantum,gard2018fast,majer2007coupling,grimsmo2013cavity,eickbusch2022fast}.
The dispersive interaction imprints \(\chi\) through a photon-number-dependent qubit
phase, making the metrological response sensitive to both cavity statistics and qubit
coherence. The same model also gives a controlled setting for comparing Markovian
decay with non-Markovian environmental memory, which we describe using an
Ornstein--Uhlenbeck bath correlation function within the quantum-state-diffusion
formalism \cite{gisin1992quantum,percival1998quantum,diosi1998non,strunz1999open,yu1999non}.

The paper is organized as follows. Sec.~II introduces the dispersive cavity-qubit model
and the QFI formalism, including the closed-system benchmark and the Markovian and
non-Markovian open-system descriptions. Sec.~III addresses the first metrological
question: which initial probe state remains most useful once open-system decay is
included. Sec.~IV studies the measurement problem by identifying a fixed observable
that approaches the Cram\'er--Rao bound over a finite range of the unknown coupling.
Finally, Sec.~V summarizes the main
conclusions and discusses extensions of the framework, including multiparameter
estimation and measurement compatibility.
 \section{Dispersive Model and QFI Formalism}

We consider a single bosonic cavity mode coupled dispersively to a two-level system. The
dynamics are described by the effective Hamiltonian
\begin{equation}
H_S=\omega_c a^\dagger a+\frac{\omega_q}{2}\sigma_z
    +\chi\,a^\dagger a\,\sigma_z ,
\label{eq:v2-dispersive-hamiltonian}
\end{equation}
where \(\omega_c\) and \(\omega_q\) denote the bare cavity and qubit frequencies,
respectively, and \(\chi\) is the dispersive coupling strength to be estimated.
Throughout this work, Eq.~\eqref{eq:v2-dispersive-hamiltonian} is taken as the effective
system Hamiltonian defining the metrological model.

The interaction term \(H_{\mathrm{int}}=\chi a^\dagger a\,\sigma_z\) has the form of a
qubit-dependent cross-Kerr interaction: the cavity frequency depends on the qubit state,
while the qubit transition frequency depends on the cavity photon number. Such effective
interactions arise naturally in dispersive cavity- and circuit-QED settings and underlie,
for example, photon-number-dependent qubit spectroscopy and quantum-nondemolition
readout
\cite{jaynes1963comparison,haroche2006exploring,blais2004cavity,blais2021circuit,wallraff2004strong,schuster2007resolving}.
Our analysis concerns estimation of the effective coupling coefficient \(\chi\),
independently of the particular microscopic mechanism used to generate this interaction.

We use the convention \(\sigma_z\ket e=\ket e\) and
\(\sigma_z\ket g=-\ket g\), and denote the cavity number operator by
\(\hat N=a^\dagger a\). The parameter \(\chi\) therefore produces opposite
photon-number-dependent phase shifts for the two qubit states. The resulting
distinguishability between nearby values of \(\chi\) forms the basis of the metrological
protocol considered throughout this work.

This structure simplifies the estimation problem and, as discussed below, explains
why the bare cavity--qubit detuning does not directly control the QFI for estimation of
\(\chi\).

\subsection{Closed-System Benchmark}

We first establish the closed-system metrological benchmark associated with the
dispersive interaction. From Eq.~\eqref{eq:v2-dispersive-hamiltonian}, with
\(\hat N=a^\dagger a\), the generator associated with the unknown dispersive coupling
\(\chi\) is
\begin{equation}
G_\chi=\frac{\partial H_S}{\partial\chi}=\hat N\sigma_z.
\label{eq:v2-generator}
\end{equation}
An important simplification of the present model is that the three contributions to
\(H_S\) commute: \([\hat N,\sigma_z]=[\hat N,\hat N\sigma_z]
=[\sigma_z,\hat N\sigma_z]=0\). The unitary evolution can consequently be factorized
as \(U_\chi(t)=e^{-i\omega_c\hat Nt}e^{-i\omega_q\sigma_zt/2}
e^{-i\chi\hat N\sigma_z t}\). Thus, the local generator of translations in \(\chi\),
\(\mathcal H_\chi=iU_\chi^\dagger(t)\partial_\chi U_\chi(t)\), reduces to
\(\mathcal H_\chi=tG_\chi=t\hat N\sigma_z\).

For an initially pure probe state \(\ket{\psi_0}\), the quantum Fisher information for
estimating \(\chi\) is therefore \(\mathcal F_\chi(t)=4\,
\mathrm{Var}_{\psi_0}(\mathcal H_\chi)=4t^2\mathrm{Var}_{\psi_0}(G_\chi)\), or
explicitly, \(\mathcal F_\chi(t)=4t^2[\langle(\hat N\sigma_z)^2\rangle-
\langle\hat N\sigma_z\rangle^2]\). Since \(\sigma_z^2=I\), this becomes
\(\mathcal F_\chi(t)=4t^2[\langle\hat N^2\rangle-
\langle\hat N\sigma_z\rangle^2]\).
For the pure probes considered in this closed-system benchmark, the QFI is therefore
directly determined by the variance of the parameter generator.
This expression provides the closed-system reference against which the open-system
dynamics will be compared below.

For a separable initial probe \(\ket{\psi_0}=\ket{\phi_c}\otimes\ket{\phi_q}\), the
expectation values factorize as \(\langle\hat N\sigma_z\rangle=
\langle\hat N\rangle\langle\sigma_z\rangle\). Consequently,
\(\mathcal F_\chi(t)=4t^2[\langle\hat N^2\rangle-
\langle\sigma_z\rangle^2\langle\hat N\rangle^2]\). Using
\(\langle\hat N^2\rangle=\mathrm{Var}(\hat N)+\langle\hat N\rangle^2\), this can be
written as
\begin{equation}
\mathcal F_\chi(t)=4t^2\left[
\mathrm{Var}(\hat N)+\left(1-\langle\sigma_z\rangle^2\right)
\langle\hat N\rangle^2\right].
\label{eq:v2-closed-product-qfi}
\end{equation}
The two terms have distinct physical interpretations. The first contribution originates
from photon-number fluctuations of the cavity field. The second originates from
coherence of the qubit between the two eigenstates of \(\sigma_z\), since
\(1-\langle\sigma_z\rangle^2\) vanishes for a \(\sigma_z\) eigenstate and is maximal
for an equal superposition of the two qubit states. Thus, even in the absence of
photon-number fluctuations, a definite nonzero photon number can provide sensitivity
to \(\chi\) when accompanied by qubit coherence. Conversely, for a qubit prepared in a
\(\sigma_z\) eigenstate, sensitivity arises entirely from photon-number fluctuations.

This distinction can be seen immediately for the single-photon product probe
\(\ket{1,\pm}=\ket{1}\otimes(\ket g\pm\ket e)/\sqrt2\). Here the cavity has
\(\langle\hat N\rangle=1\) and \(\mathrm{Var}(\hat N)=0\), while the
qubit satisfies \(\langle\sigma_z\rangle=0\). Therefore,
\(\mathcal F_\chi^{\ket{1,\pm}}(t)=4t^2\). In contrast, for \(\ket{1,g}\) or
\(\ket{1,e}\), \(\langle\sigma_z\rangle^2=1\), and hence
\(\mathcal F_\chi^{\ket{1,g}}(t)=\mathcal F_\chi^{\ket{1,e}}(t)=0\).
The physical origin of this difference is transparent. The two components of
\(\ket{1,+}\) acquire opposite \(\chi\)-dependent phases under the dispersive
interaction, producing a relative phase proportional to \(2\chi t\). A qubit energy
eigenstate acquires only an overall \(\chi\)-dependent phase and therefore carries no
observable information about \(\chi\).

The optimality condition can also be obtained directly from the spectrum of \(G_\chi\).
In the single-photon cavity truncation \(\mathcal H_c=\operatorname{span}
\{\ket0,\ket1\}\), the eigenvalues of \(G_\chi=\hat N\sigma_z\) are
\(\{0,0,+1,-1\}\). For any Hermitian
generator with minimum and maximum eigenvalues \(g_{\min}\) and \(g_{\max}\), the
largest possible variance is
\begin{equation}
\mathrm{Var}(G)_{\max}=\frac{(g_{\max}-g_{\min})^2}{4},
\end{equation}
which is attained by an equal-weight coherent superposition of eigenstates belonging
to the two extremal eigenvalues. In the present case, \(g_{\max}=+1\) and
\(g_{\min}=-1\), and therefore \(\mathrm{Var}(G_\chi)_{\max}=1\), giving the
closed-system bound
\begin{equation}
\mathcal F_{\chi,\max}^{\mathrm{closed}}(t)=4t^2.
\label{eq:v2-closed-bound}
\end{equation}
A corresponding family of optimal probes is \(\ket{\psi_{\mathrm{opt}}}=
(\ket{1,g}+e^{i\varphi}\ket{1,e})/\sqrt2\), where the relative phase \(\varphi\) is
arbitrary. In particular, the state
\(\ket{1,+}\) used below belongs to this optimal family.

It is useful to contrast this state with the cavity--qubit product superposition
\(\ket{+,+}=(\ket0+\ket1)\otimes(\ket g+\ket e)/2\).
For this state, \(\langle G_\chi\rangle=0\) and
\(\langle G_\chi^2\rangle=1/2\), so that
\begin{equation}
\mathcal F_\chi^{\ket{+,+}}(t)=2t^2.
\end{equation}
The reduction relative to \(\ket{1,+}\) arises because half of the probability
amplitude occupies the vacuum sector, for which \(G_\chi\ket{0,q}=0\). The vacuum
components therefore do not accumulate a dispersive phase proportional to \(\chi\).

The role of the bare frequencies is also transparent. Writing
\begin{equation}
U_\chi(t)=U_0(t)e^{-i\chi tG_\chi},\qquad
U_0(t)=e^{-i\omega_c\hat Nt}e^{-i\omega_q\sigma_z t/2},
\end{equation}
we see that the parameter-independent unitary \(U_0(t)\) cannot alter the QFI for
estimating \(\chi\). Equivalently, neither \(\omega_c\) nor \(\omega_q\) appears in
Eqs.~\eqref{eq:v2-closed-product-qfi}--\eqref{eq:v2-closed-bound}. Thus, within the
closed effective model, the QFI is exactly independent of the cavity--qubit detuning
\begin{equation}
\Delta=\omega_c-\omega_q.
\end{equation}

This closed-system analysis establishes both a benchmark for probe optimality and the
two principal resources for estimating \(\chi\): photon-number statistics and qubit
coherence. We next examine how these features are modified by cavity loss and
environmental memory.

\subsection{Markovian dynamics and stochastic quantum Fisher information}

We model cavity photon loss through the coupling operator
$L=\sqrt{\Gamma}\,a$. In the Markovian limit, the unconditional
reduced state of the cavity--qubit system obeys the Lindblad master
equation
\begin{equation}
\dot{\rho}_{\chi}
=
-i[H_S,\rho_{\chi}]
+
\Gamma
\left(
a\rho_{\chi}a^\dagger
-\frac{1}{2}\{a^\dagger a,\rho_{\chi}\}
\right),
\label{eq:lindblad}
\end{equation}
where $\Gamma$ denotes the cavity photon-number decay rate for the Lindblad convention
used here; hence \(\Gamma^{-1}\) is the cavity photon-number decay time. Because
\(\chi\) has dimensions of frequency, \(F_\chi\) has dimensions of inverse frequency
squared. The quantum Fisher
information associated with the unconditional state is evaluated from
\begin{equation}
F_{\chi}^{(\rho)}
=
2\sum_{m,n}
\frac{
|\langle m|\partial_{\chi}\rho_{\chi}|n\rangle|^2
}{
\lambda_m+\lambda_n
},
\end{equation}
where
$\rho_{\chi}=\sum_m\lambda_m|m\rangle\langle m|$,
and terms for which $\lambda_m+\lambda_n=0$ are omitted.
This quantity therefore characterizes the information about $\chi$
available from the unconditional reduced state after averaging over
the environmental degrees of freedom.

The independence from the bare cavity--qubit detuning extends to this
Markovian model. Transforming to the interaction picture generated by
\begin{equation}
H_0=\omega_c\hat{N}+\frac{\omega_q}{2}\sigma_z
\end{equation}
removes the free Hamiltonian terms. Since
\begin{equation}
e^{iH_0t}a e^{-iH_0t}=e^{-i\omega_c t}a,
\end{equation}
the phase acquired by the collapse operator cancels from the
dissipator,
$\mathcal{D}[e^{-i\omega_c t}a]=\mathcal{D}[a]$.
The interaction-picture equation is therefore
\begin{equation}
\dot{\widetilde{\rho}}_{\chi}
=
-i[\chi\hat{N}\sigma_z,\widetilde{\rho}_{\chi}]
+
\Gamma
\left(
a\widetilde{\rho}_{\chi}a^\dagger
-\frac{1}{2}
\{a^\dagger a,\widetilde{\rho}_{\chi}\}
\right),
\end{equation}
which contains neither $\omega_c$ nor $\omega_q$. The transformation
back to the Schr\"odinger picture is independent of $\chi$ and hence
does not change the QFI. Therefore, the QFI for estimating $\chi$ is
also independent of the cavity--qubit detuning in the Markovian
dynamics considered here.

\begin{figure}[t]
    \centering
    \includegraphics[width=\columnwidth]{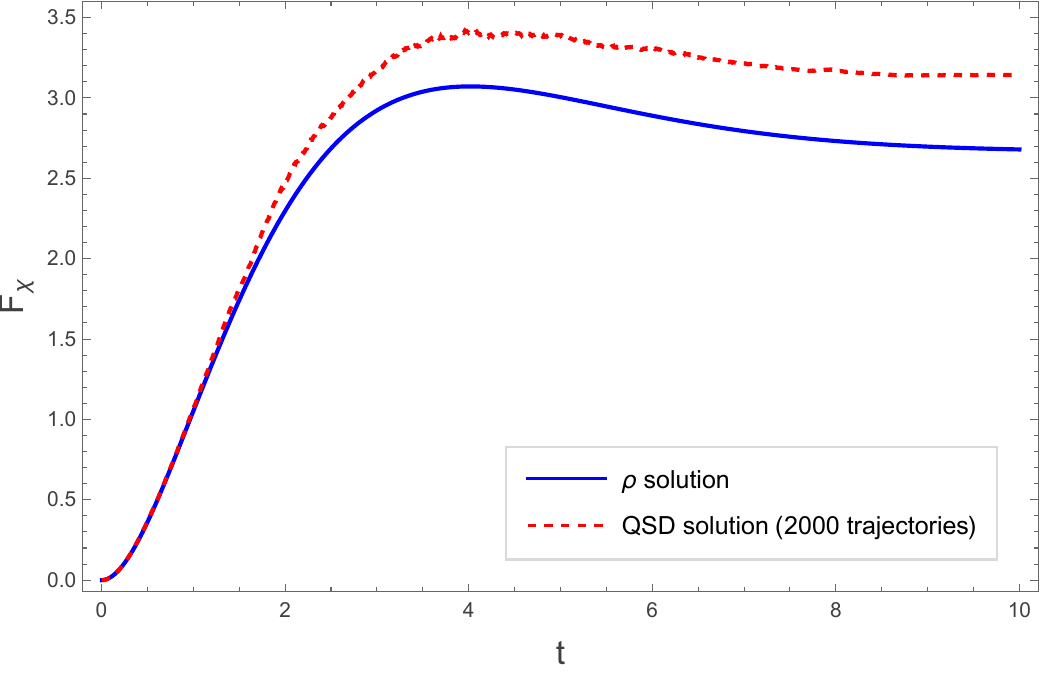}
    \caption{Time evolution of the unconditional density-matrix QFI (blue solid
    curve) and the ensemble-averaged SQFI obtained from QSD simulations with
    $N_{\mathrm{traj}}=2000$ trajectories (red dashed curve). The density-matrix
    result is obtained by solving the Markovian Lindblad master equation, and
    the trajectory simulations use an integration time step $\Delta t=0.01$.
    Both curves correspond to the initial state $\ket{+,+}$, with
    $\omega_c=\omega_q=1$, $\Gamma=1$, and $\chi=0.2$.}
    \label{fig:sqfi_validation}
\end{figure}

A complementary description is provided by the
quantum-state-diffusion (QSD) unraveling of the same master equation.
For the analytical calculation, we begin with the linear QSD equation
\begin{equation}
\partial_t|\psi_t\rangle
=
\left[
-iH_S-\frac{\Gamma}{2}a^\dagger a
+\sqrt{\Gamma}\,a z_t^*
\right]|\psi_t\rangle,
\end{equation}
where $z_t$ is a complex Gaussian white-noise process satisfying
\begin{equation}
\mathcal{M}[z_t^*]=0,
\qquad
\mathcal{M}[z_t^*z_s^*]=0,
\qquad
\mathcal{M}[z_t^*z_s]=\delta(t-s).
\end{equation}
The linear form is particularly convenient analytically because the
trajectory amplitudes obey a closed recursive hierarchy in the joint
photon--qubit basis. The corresponding coefficient-level derivation is
given in Appendix~A.

For a normalized pure trajectory, the quantum Fisher information for
estimating $\chi$ is
\begin{equation}
F_{\chi}[z_t^*]
=
4\left[
\langle\partial_{\chi}\widetilde{\psi}_t|
\partial_{\chi}\widetilde{\psi}_t\rangle
-
\left|
\langle\widetilde{\psi}_t|
\partial_{\chi}\widetilde{\psi}_t\rangle
\right|^2
\right],
\end{equation}
where $|\widetilde{\psi}_t\rangle$ denotes the normalized trajectory
associated with a particular noise realization. We refer to this
trajectory quantity as the stochastic quantum Fisher information
(SQFI). Its ensemble average is
\begin{equation}
\overline{F}_{\chi}^{\mathrm{SQFI}}(t)
=
\mathcal{M}[F_{\chi}[z_t^*]].
\end{equation}

For numerical trajectory simulations, normalized QSD trajectories are
generated using the corresponding nonlinear QSD equation obtained
through the Girsanov transformation. In Stratonovich form, this can be
written as
\begin{align}
\partial_t|\widetilde{\psi}_t\rangle
={}&
-iH_S|\widetilde{\psi}_t\rangle
+
(L-\langle L\rangle_t)|\widetilde{\psi}_t\rangle
\circ
\left(z_t^*+\langle L^\dagger\rangle_t\right)
\nonumber\\
&-\frac{1}{2}
\left(
L^\dagger L-\langle L^\dagger L\rangle_t
\right)
|\widetilde{\psi}_t\rangle ,
\end{align}
with $L=\sqrt{\Gamma}\,a$. This nonlinear representation preserves
the normalization of individual trajectories and provides the
physical trajectory measure used in the numerical ensemble average.
The linear and nonlinear QSD equations constitute equivalent
unravelings of the same Markovian reduced dynamics, with the Girsanov
transformation relating the corresponding stochastic measures.

For the balanced cavity--qubit superposition considered in the
analytical calculation, the stochastic average can be evaluated
explicitly. Upon measuring time in units of the cavity decay rate, one
obtains
\begin{equation}
\begin{split}
    \overline{\mathcal{F}}_\chi(t)
    &=
    4\left[1-e^{-t}(1+t)\right]
    -\int_0^t ds\,s^2e^{-s}\sin^2(\chi s)
    \\
    &\quad
    -\int_0^t ds\int_0^t ds'\,
    {s'}^2e^{-(s+s')}\sin^2\!\left[\chi(s-s')\right].
\end{split}
\end{equation}
The detailed derivation from the trajectory coefficients is presented
in Appendix~A. This expression also makes the frequency independence
particularly transparent: the final analytical SQFI contains $\chi$,
but neither $\omega_c$ nor $\omega_q$. Thus, the analytical trajectory
calculation independently confirms the absence of cavity--qubit
detuning dependence in the Markovian SQFI.

For the numerical stochastic calculations, we use Monte Carlo simulations
of the QSD dynamics with a fixed integration time step $\Delta t=0.01$
and $N_{\mathrm{traj}}=2000$ trajectories. Figure~\ref{fig:sqfi_validation}
shows the resulting ensemble-averaged SQFI, which initially grows,
reaches a finite-time maximum, and then decreases toward a nonzero
long-time value. The finite trajectory ensemble introduces Monte Carlo
sampling fluctuations into the numerical curve.

For comparison, we also evaluate the QFI directly from the
unconditional density matrix obtained by solving the Lindblad master
equation. The resulting evolution is shown in Fig.~\ref{fig:sqfi_validation}.
The density-matrix QFI exhibits the same qualitative behavior as the
ensemble-averaged SQFI: it initially increases, reaches a finite-time
maximum, and subsequently relaxes toward a nonzero long-time value.
For the parameters considered here, the density-matrix QFI reaches a
maximum of approximately $F_\chi^{(\rho)}\simeq 3.072$ at
$t\simeq4.019$.

At long times, however, the ensemble-averaged SQFI approaches a
slightly larger value than the QFI obtained from the unconditional
density matrix. This difference is expected and reflects the distinct
information content of the two quantities. The density-matrix QFI is
evaluated only after averaging over the environmental noise,
\begin{equation}
\rho_\chi(t)
=
\mathcal{M}_{z}
\left[
|\widetilde{\psi}_{\chi,z}(t)\rangle
\langle\widetilde{\psi}_{\chi,z}(t)|
\right],
\end{equation}
whereas the SQFI is first evaluated for each conditioned pure-state
trajectory and then averaged,
\begin{equation}
\overline{F}_{\chi}^{\mathrm{SQFI}}(t)
=
\mathcal{M}_{z}
\left[
F_\chi\!\left(
|\widetilde{\psi}_{\chi,z}(t)\rangle
\right)
\right].
\end{equation}
Since the QFI is convex under parameter-independent mixing, these two operations do
not in general commute, and one has
\begin{equation}
F_\chi^{(\rho)}(t)
\leq
\overline{F}_{\chi}^{\mathrm{SQFI}}(t).
\end{equation}
This inequality follows from the convex-roof characterization of the QFI and its
convexity under parameter-independent mixing~\cite{TothPetz2013,TothFrowis2022}.
Thus, averaging over the stochastic trajectories at the density-matrix level can
discard information about $\chi$ that remains accessible when the individual
conditioned trajectories are retained. The slightly higher long-time SQFI observed
here is therefore consistent with the expected relation between the conditioned and
unconditional descriptions, rather than being a numerical discrepancy.

\subsection{Non-Markovian Quantum-State-Diffusion Dynamics}

To investigate the effect of finite environmental memory, we replace the white-noise
reservoir by an exponentially correlated Gaussian environment with correlation function
\begin{equation}
\alpha(t,s)=\frac{\gamma}{2}
e^{-\gamma|t-s|-i\Omega(t-s)},
\label{eq:v2-ou-correlation}
\end{equation}
Here \(\gamma\) determines the inverse environmental correlation time,
\begin{equation}
\tau_B\sim\frac{1}{\gamma},
\end{equation}
such that smaller \(\gamma\) corresponds to longer-lived bath correlations, while
larger \(\gamma\) drives the reservoir toward the short-memory limit. The parameter
\(\Omega>0\) specifies the oscillatory component of the correlation function and
therefore determines the spectral position of the structured reservoir. The overall
system--environment coupling strength is controlled independently by \(\Gamma\).

Non-Markovian QSD has also been applied to the estimation of entanglement dynamics
in two-qubit systems coupled to a common environment, illustrating the utility of
diffusive trajectories for tracking quantum correlations in the presence of
environmental memory \cite{CornJingYu2016}.
Within the non-Markovian quantum-state-diffusion formalism
\cite{diosi1998non,strunz1999open,yu1999non}, the stochastic state satisfies
\begin{equation}
\frac{d}{dt}\ket{\psi_t}
=\left[-iH_S+\sqrt{\Gamma}\,a z_t^*-\sqrt{\Gamma} a^\dagger\overline O(t)\right]\ket{\psi_t},
\label{eq:v2-nm-qsd-main}
\end{equation}
with colored complex Gaussian noise obeying
\begin{equation}
\mathcal M[z_t]=0,\qquad \mathcal M[z_tz_s]=0,\qquad
\mathcal M[z_t^*z_s]=\alpha(t,s).
\label{eq:v2-nm-noise}
\end{equation}

The functional derivative of the stochastic state is written in terms of the
\(O\)-operator as
\begin{equation}
\frac{\delta\ket{\psi_t}}{\delta z_s^*}
=O(t,s)\ket{\psi_t},
\end{equation}
This convention isolates the coupling strength \(\Gamma\) from the bath
correlation function, which characterizes the temporal structure of the environmental
noise. The corresponding memory operator is
\begin{equation}
\overline O(t)=\int_0^t ds\,\alpha(t,s)O(t,s).
\label{eq:v2-nm-obar-definition}
\end{equation}
For the dispersive Hamiltonian considered here, the \(O\)-operator closes within the
two-dimensional operator basis
\begin{equation}
O(t,s)=f_1(t,s)a+f_2(t,s)a\sigma_z,
\label{eq:v2-nm-o-operator}
\end{equation}
with boundary conditions \(f_1(t,t)=\sqrt{\Gamma}\) and \(f_2(t,t)=0\). Introducing
\begin{equation}
F_j(t)=\int_0^t ds\,\alpha(t,s)f_j(t,s),\qquad j=1,2,
\end{equation}
the memory operator takes the form
\begin{equation}
\overline O(t)=F_1(t)a+F_2(t)a\sigma_z.
\label{eq:v2-main-bar-o}
\end{equation}
The resulting time-local reduced master equation is
\begin{equation}
\dot\rho_\chi=-i[H_S,\rho_\chi]
+\sqrt{\Gamma}[a,\rho_\chi\overline O^\dagger(t)]
+\sqrt{\Gamma}[\overline O(t)\rho_\chi,a^\dagger].
\label{eq:v2-nm-master-main}
\end{equation}
The first term in Eq.~\eqref{eq:v2-nm-master-main} describes the coherent dispersive
evolution, while the remaining terms contain the finite-time memory response of the
bath through \(F_1(t)\) and \(F_2(t)\).

The consistency condition for the \(O\)-operator yields
\begin{align}
\dot f_1&=i(\omega_c f_1+\chi f_2)
+\sqrt{\Gamma}(F_1f_1+F_2f_2),
\label{eq:v2-f1-evolution}\\
\dot f_2&=i(\omega_c f_2+\chi f_1)
+\sqrt{\Gamma}(F_1f_2+F_2f_1).
\label{eq:v2-f2-evolution}
\end{align}
It is convenient to introduce the combinations \(X_\pm(t)=F_1(t)\pm F_2(t)\).
For the exponential correlation function in Eq.~\eqref{eq:v2-ou-correlation}, these
quantities satisfy the decoupled Riccati equations
\begin{align}
\dot X_\pm&=\frac{\gamma}{2}+\xi_\pm X_\pm+\sqrt{\Gamma} X_\pm^2,
\label{eq:v2-main-Xpm-solution}\\
\xi_\pm&=-\gamma+i(\omega_c-\Omega\pm\chi).
\label{eq:v2-xi-pm}
\end{align}
The two functions \(X_\pm\) describe the bath response associated with the two
qubit-conditioned cavity frequencies \(\omega_c\pm\chi\). The original memory
coefficients are recovered from
\begin{equation}
F_1(t)=\frac{X_+(t)+X_-(t)}{2},\qquad
F_2(t)=\frac{X_+(t)-X_-(t)}{2}.
\label{eq:v2-main-F1F2}
\end{equation}
These coefficients determine the time-dependent modification of the cavity damping
induced by environmental memory.

Equation~\eqref{eq:v2-xi-pm} reveals two features relevant to the choice of numerical
parameters. First, \(\omega_q\) is absent from the non-Markovian memory coefficients.
Consequently, varying \(\omega_q\) at fixed \(\omega_c\) changes the bare cavity--qubit
detuning
\begin{equation}
\Delta=\omega_c-\omega_q
\end{equation}
without modifying \(X_\pm\), \(F_1\), or \(F_2\). Together with the commuting structure
of \(H_S\), this shows that the bare cavity--qubit detuning does not provide an
independent control parameter for the \(\chi\)-QFI in the present model. We have also
verified this behavior numerically by varying \(\omega_q\) at fixed \(\omega_c\).

Second, unlike \(\omega_q\), the cavity frequency enters the non-Markovian memory
coefficients through the combination
\begin{equation}
S=\omega_c-\Omega.
\end{equation}
Thus, changing \(\omega_c\) while holding \(\Omega\) fixed does not constitute a pure
variation of the cavity--qubit detuning: it simultaneously changes the position of the
structured reservoir relative to the qubit-conditioned cavity frequencies. The roles
of the cavity--qubit detuning and of \(S\) must therefore be distinguished.

\begin{figure}[t]
    \centering
    \includegraphics[width=\columnwidth]{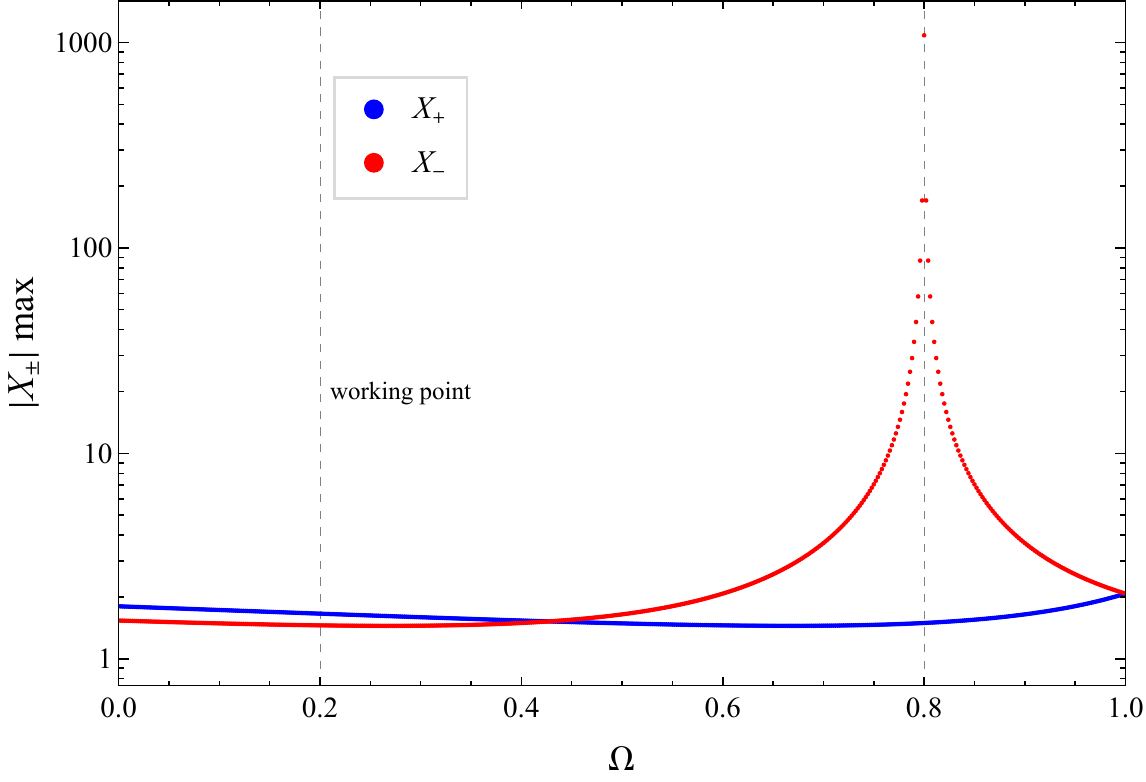}
    \caption{Maximum magnitude of the non-Markovian response functions
    \(X_+(t)\) and \(X_-(t)\) over the evolution interval as a function of
    \(\Omega/\Gamma\), for \(\omega_c/\Gamma=1\), \(\gamma/\Gamma=0.6\), and
    \(\chi/\Gamma=0.2\). The \(X_-\) branch exhibits a sharp pole-like enhancement
    near \(\Omega/\Gamma\simeq0.8\), while \(X_+\) remains bounded over the displayed
    range. The dashed vertical line marks the working point \(\Omega/\Gamma=0.2\)
    used in the subsequent numerical calculations.}
    \label{fig:v2-nm-numerical-stability}
\end{figure}

The analytical solution of Eq.~\eqref{eq:v2-main-Xpm-solution}, derived in Appendix~B,
can be written as
\begin{equation}
X_\pm(t)=\delta_\pm\left(
\frac{2}{1-c_\pm e^{2\sqrt{\Gamma}\delta_\pm t}}-1\right)
-\frac{\xi_\pm}{2\sqrt{\Gamma}},
\label{eq:v2-Xpm-closed-form}
\end{equation}
where
\begin{equation}
\delta_\pm^2=\frac{\xi_\pm^2}{4\Gamma}-\frac{\gamma}{2},
\qquad
c_\pm=\frac{\xi_\pm/(2\sqrt{\Gamma})-\delta_\pm}
{\xi_\pm/(2\sqrt{\Gamma})+\delta_\pm}.
\label{eq:v2-delta-c-pm}
\end{equation}
The denominator
\begin{equation}
D_\pm(t)=1-c_\pm e^{2\sqrt{\Gamma}\delta_\pm t}
\end{equation}
shows that the closed-form memory coefficients can become strongly enhanced when
\(D_\pm(t)\) approaches zero. Since
\begin{equation}
\xi_+=-\gamma+i(S+\chi),\qquad \xi_-=-\gamma+i(S-\chi),
\end{equation}
the two branches exhibit pole-like enhancements around
\begin{equation}
S=-\chi\quad (X_+),\qquad S=+\chi\quad (X_-).
\end{equation}
For \(\chi/\Gamma=0.2\), these occur at \(S/\Gamma=-0.2\) and
\(S/\Gamma=+0.2\), respectively. For the fixed value \(\omega_c/\Gamma=1\),
these locations correspond to \(\Omega/\Gamma=1.2\) and
\(\Omega/\Gamma=0.8\). Figure~\ref{fig:v2-nm-numerical-stability} shows the
maximum magnitude of \(X_+(t)\) and \(X_-(t)\) over the evolution interval as
a function of \(\Omega/\Gamma\). The \(X_-\) enhancement lies within the displayed
range, whereas the corresponding \(X_+\) enhancement lies outside it. Away from
these neighborhoods, both response functions remain bounded over the investigated
parameter range.

The precise widths of the excluded regions depend on \(\gamma\) and on the
adopted coefficient cutoff, but remain localized around \(S=\pm\chi\).
We denote the corresponding excluded neighborhoods in the dimensionless variable
\(S/\Gamma\) by
\begin{equation}
\mathcal E_\pm(\gamma,F_{\mathrm{cut}})
\quad\text{centered at}\quad S/\Gamma=\mp\chi/\Gamma,
\end{equation}
where the subscripts identify the \(X_\pm\) branches. The cutoff-dependent
excluded region associated with \(X_-\) in the
\((\gamma,\omega_c-\Omega)\) plane is presented in Appendix~B and shown in
Fig.~\ref{fig:v2-app-nm-parameter-bounds}.

For the numerical results reported below, we require the working parameters to
remain outside both cutoff-dependent excluded neighborhoods,
\begin{equation}
S/\Gamma\notin\mathcal E_+(\gamma,F_{\mathrm{cut}})
\cup\mathcal E_-(\gamma,F_{\mathrm{cut}}).
\label{eq:v2-stability-condition}
\end{equation}
This restriction is a numerical stability criterion for the parameter range and
evolution times considered here, rather than a fundamental physical constraint
on the dispersive model.

An important consistency check is obtained in the short-memory limit. For sufficiently
large \(\gamma\), the dominant terms in Eq.~\eqref{eq:v2-main-Xpm-solution} are
\begin{equation}
\dot X_\pm\simeq\frac{\gamma}{2}-\gamma X_\pm,
\end{equation}
so that, after the initial bath transient, \(X_\pm\to 1/2\). Consequently,
\(F_1(t)\to 1/2\) and \(F_2(t)\to0\). Substituting these limits into
Eq.~\eqref{eq:v2-nm-master-main} gives
\begin{equation}
\dot\rho_\chi\longrightarrow-i[H_S,\rho_\chi]
+\Gamma\left[a\rho_\chi a^\dagger
-\frac{1}{2}\left\{a^\dagger a,\rho_\chi\right\}\right],
\end{equation}
which reproduces the Markovian master equation in Eq.~\eqref{eq:lindblad}.
Thus, at fixed coupling strength \(\Gamma\), the parameter \(\gamma\) provides a
continuous interpolation between finite-memory dynamics and the Markovian limit. This
correspondence allows changes in the metrological sensitivity to be associated
specifically with environmental memory rather than with a simultaneous change of the
overall system--bath coupling strength.

Unless explicitly varied, we therefore use the common parameter set
\begin{equation}
\begin{gathered}
\omega_c/\Gamma=\omega_q/\Gamma=1,\quad \gamma/\Gamma=0.6,\\
\Omega/\Gamma=0.2,\quad \chi/\Gamma=0.2
\end{gathered}
\label{eq:v2-common-parameters}
\end{equation}
throughout the numerical results. We use the dimensionless time
\(\tau=\Gamma t\), with \(\Gamma\) setting the frequency and time scales; for example,
\(\Gamma t\simeq4\) corresponds to approximately four cavity photon-number decay times.
Accordingly, the dimensionless QFI reported under this normalization is
\(\Gamma^2F_\chi\). The choice \(\omega_c/\Gamma=\omega_q/\Gamma=1\) fixes a convenient
common frequency ratio; it is not required by the metrological protocol, since the
cavity--qubit detuning does not affect the QFI for \(\chi\) in the present model.
With the present convention for the bath correlation function,
\((\omega_c-\Omega)/\Gamma=0.8\), placing the chosen non-Markovian parameters well
outside both pole-like regions centered at
\((\omega_c-\Omega)/\Gamma=\pm\chi/\Gamma=\pm0.2\). This parameter
set therefore provides a consistent and numerically stable reference point for
comparing probe states and Markovian and non-Markovian dynamics in the following
sections.
 ﻿\section{Open-System Probe Selection}
\label{sec:probe_selection}

The closed-system analysis of Sec.~II establishes a natural benchmark for probe
selection. In the single-photon cavity truncation, the generator
\(G_\chi=\sigma_z\hat N\) has extremal eigenvalues \(\pm1\), and the state
\(\ket{1,+}\) belongs to the family of probes that maximize its variance, yielding
\(\mathcal F_\chi=4t^2\) in the absence of environmental coupling. By comparison, the
product superposition \(\ket{+,+}=(\ket0+\ket1)(\ket g+\ket e)/2\) gives
\(\mathcal F_\chi=2t^2\), since half of its population lies in the vacuum sector and
therefore does not acquire a \(\chi\)-dependent dispersive phase. The central question
in the open system is consequently whether this closed-system hierarchy survives
dissipation, and how it is modified when the environment possesses a finite memory
time.

\begin{figure*}[t]
\centering
\begin{minipage}{0.96\textwidth}
\begin{minipage}[t]{0.48\textwidth}
    \centering
    \includegraphics[width=\linewidth]{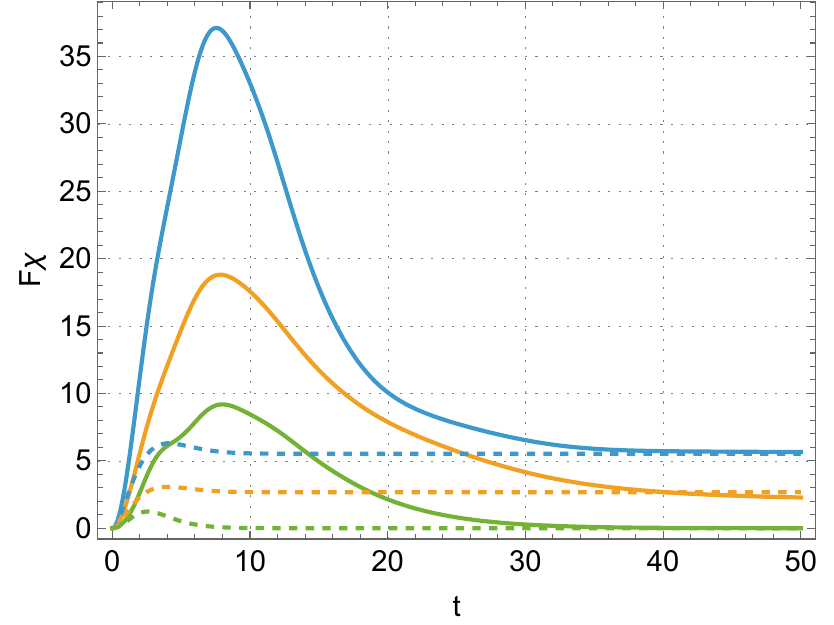}
    \vspace{0.25em}

    \textbf{(a)} QFI evolution
\end{minipage}
\hfill
\begin{minipage}[t]{0.48\textwidth}
    \centering
    \includegraphics[width=\linewidth]{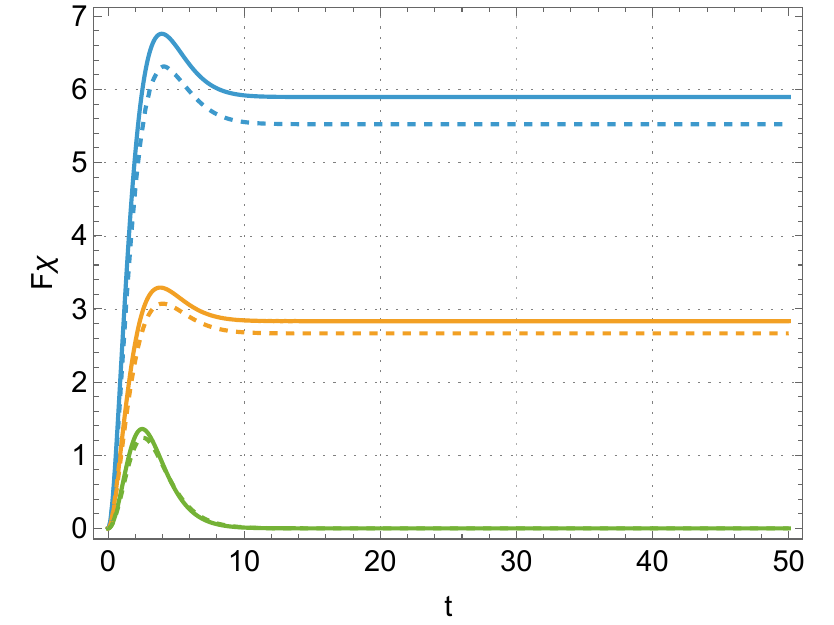}
    \vspace{0.25em}

    \textbf{(b)} Short-memory limit
\end{minipage}
\caption{Open-system QFI for representative probes. Panel~(a) shows the QFI
evolution for \(\ket{1,+}\) (blue), \(\ket{+,+}\) (orange), and the entangled state
\(\ket{\psi(7\pi/20)}=\cos(7\pi/20)\ket{0,e}+\sin(7\pi/20)\ket{1,g}\) (green), using
\(\omega_c/\Gamma=\omega_q/\Gamma=1\), \(\gamma/\Gamma=0.6\),
\(\Omega/\Gamma=0.2\), and \(\chi/\Gamma=0.2\).
Solid curves denote non-Markovian evolution and dashed curves denote Markovian
evolution. Panel~(b) shows the corresponding evolution for the shorter-memory case
\(\gamma/\Gamma=10\), with all other parameters unchanged. As \(\gamma\) increases, the non-Markovian curves
approach their Markovian counterparts, consistent with the short-memory limit derived
in Sec.~II C.}
\label{fig:v2-qfi-states}
\end{minipage}
\end{figure*}

Figure~\ref{fig:v2-qfi-states}(a) compares the QFI evolution for representative initial
probes under Markovian and non-Markovian dynamics. Among the states shown,
\(\ket{1,+}\) retains the largest QFI in both environments. This behavior is consistent
with its closed-system structure: all of the cavity population initially occupies the
single-photon sector, while the qubit superposition provides coherence between the two
extremal eigenvalue sectors of \(G_\chi\). Cavity loss nevertheless strongly suppresses
the attainable sensitivity in the Markovian regime. The non-Markovian dynamics, in
contrast, produce a substantially larger finite-time QFI and extend the time interval
over which useful sensitivity is retained. Thus, the advantage of \(\ket{1,+}\) is not
eliminated by the open-system dynamics considered here; rather, the magnitude and
lifetime of its metrological sensitivity depend strongly on the structure of the
environment.

The role of environmental memory can be examined more directly through the bath
correlation parameter \(\gamma\). For the Ornstein--Uhlenbeck correlation function
introduced in Sec.~II, the bath correlation time scales as \(\tau_B\sim\gamma^{-1}\).
Increasing \(\gamma\) therefore shortens the memory time and continuously drives the
non-Markovian dynamics toward the Markovian limit, as illustrated in
Fig.~\ref{fig:v2-qfi-states}(b). For the shorter-memory environment, the non-Markovian
curves approach their corresponding Markovian results, consistent with the analytical
limit \(F_1(t)\to1/2\) and \(F_2(t)\to0\) derived in Sec.~II. The comparison identifies
the enhancement observed for smaller \(\gamma\) as a finite-memory effect:
environmental memory can preserve or recover part of the \(\chi\)-dependent
distinguishability that is rapidly lost under Markovian photon decay. Importantly, this
enhancement does not represent a new closed-system metrological resource; it reflects a
modification of how the information initially encoded in the probe is retained during
the open evolution.

\begin{figure*}[t]
\centering
\begin{minipage}{\textwidth}
    \centering
    \includegraphics[height=0.44\textwidth]{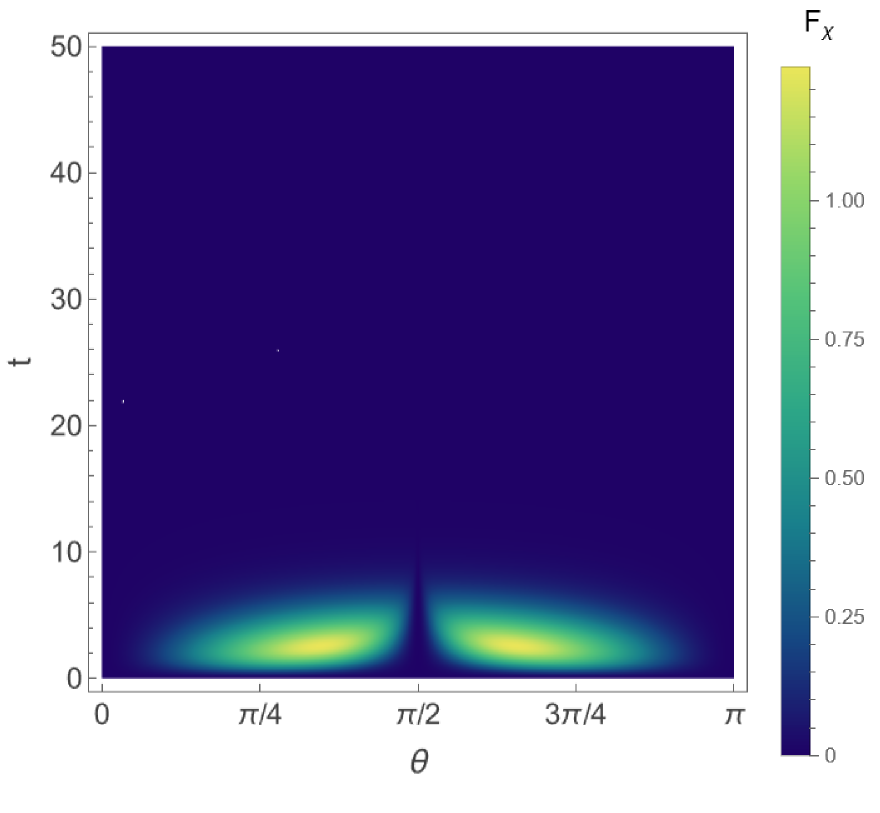}
    \hfill
    \includegraphics[height=0.44\textwidth]{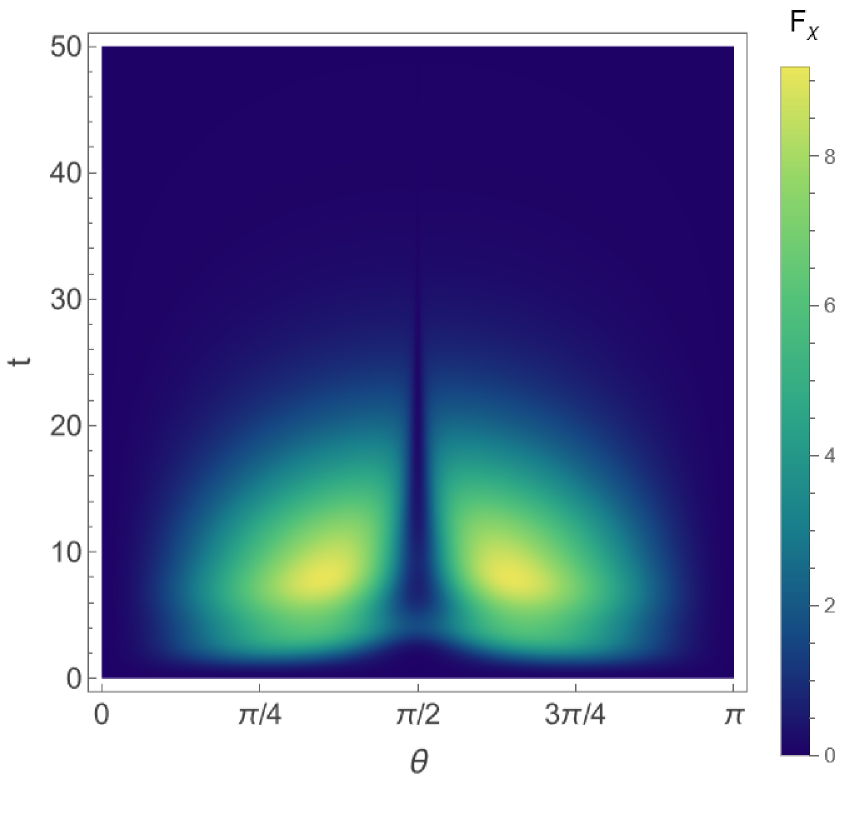}
    \par\vspace{0.25em}
    \makebox[0.47610\textwidth]{\textbf{(a)} Markovian}\hfill
    \makebox[0.46031\textwidth]{\textbf{(b)} Non-Markovian}
    \caption{QFI for the family
    \(\ket{\psi(\theta)}=\cos\theta\ket{0,e}+\sin\theta\ket{1,g}\) as a function
    of \(\theta\) and evolution time \(t\), for (a) Markovian and (b) non-Markovian
    dynamics. The reference parameters are
    \(\omega_c/\Gamma=\omega_q/\Gamma=1\), \(\chi/\Gamma=0.2\), and, in panel~(b),
    \(\gamma/\Gamma=0.6\) and \(\Omega/\Gamma=0.2\).
    Both panels span \(0\leq\theta\leq\pi\) and \(0\leq t\leq50\).
    Markovian sensitivity is concentrated at early times, whereas finite memory
    substantially extends its persistence. The high-QFI regions lie away from
    the maximally entangled states at \(\theta=\pi/4\) and \(3\pi/4\).
    Color represents \(F_\chi\), with a separate numerical color scale for each
    panel.}
    \label{fig:entangled_density}
\end{minipage}
\end{figure*}

The comparison of product probes already suggests that the relevant resource is not
simply the presence of superposition, but its distribution among the eigenspaces of
\(G_\chi\) and its subsequent robustness to cavity loss. To examine more systematically
how entanglement influences this response, we consider the family
\begin{equation}
\ket{\psi(\theta)}=\cos\theta\ket{0,e}+\sin\theta\ket{1,g}.
\label{eq:v2-entangled-probe-family}
\end{equation}
Figure~\ref{fig:entangled_density} shows the QFI as a function of both \(\theta\) and
interrogation time for Markovian and non-Markovian dynamics. Over
\(0\leq\theta\leq\pi\), the equal-weight, maximally entangled states occur at
\(\theta=\pi/4\) and \(3\pi/4\). In both environments, the high-QFI regions are
centered away from these values: maximal entanglement does not coincide with maximal
metrological usefulness for this family. The plots also resolve the temporal
structure of the response. Markovian cavity loss confines appreciable sensitivity
primarily to early times, whereas finite environmental memory extends it over a
substantially longer interval. Together with the magnitude enhancement seen in
Fig.~\ref{fig:v2-qfi-states}, this shows how memory can improve both the strength and
persistence of the response while preserving its qualitative dependence on \(\theta\).

Changing \(\theta\) changes the weights of the vacuum and single-photon sectors,
and hence both the initial dispersive response and its susceptibility to photon
loss. Entanglement alone is therefore not a sufficient criterion for probe selection.
The relevant features are the distribution of amplitude and coherence among sectors
with different dispersive responses and the robustness of those components under
open-system evolution. These density plots do not select a single value of
\(\theta\) as optimal at all interrogation times.

A further effect arises for probes that are insensitive to \(\chi\) in the
Markovian limit. The product state \(\ket{1,g}\) is an eigenstate of the closed-system
Hamiltonian and acquires only an overall \(\chi\)-dependent phase. Its generator
variance and closed-system QFI therefore vanish. Under the Markovian cavity-loss
model, the reduced state is a mixture of \(\ket{1,g}\) and \(\ket{0,g}\) with
\(\chi\)-independent populations. No \(\chi\)-dependent coherence is generated,
so the reduced density operator remains independent of \(\chi\) and \(F_\chi=0\).

In the non-Markovian dynamics, however, the response functions
\(X_\pm=F_1\pm F_2\) depend on the qubit-conditioned cavity frequencies
\(\omega_c\pm\chi\), as derived in Sec.~II. Consequently, the memory-dependent
reduced evolution can encode \(\chi\) in the populations even for this initial
state with vanishing generator variance. This memory-induced activation of parameter
sensitivity does not require the generation of coherence: a dependence of the
surviving single-photon population on \(\chi\) suffices to make the reduced states
distinguishable.

Figure~\ref{fig:memory_activated_qfi} illustrates this effect through a scan of the
inverse bath memory time \(\gamma\) and evolution time for \(\ket{1,g}\).
Finite QFI develops in suitable regimes with sufficiently long bath memory, with
pronounced temporal structure at smaller \(\gamma\); sensitivity is suppressed
toward the short-memory side of the scan. This behavior is conditional on the bath
parameters, rather than a universal consequence of non-Markovian evolution.
For the reference bath parameters of Fig.~\ref{fig:entangled_density}(b), the
\(\theta=\pi/2\) line, which corresponds to \(\ket{1,g}\), remains strongly
suppressed. The broader scan in Fig.~\ref{fig:memory_activated_qfi} shows that the
same state can acquire finite QFI in other finite-memory regimes, so the two
figures describe complementary parameter ranges.

\clearpage
\begin{center}
\begin{minipage}{\linewidth}
    \centering
    \includegraphics[width=\columnwidth]{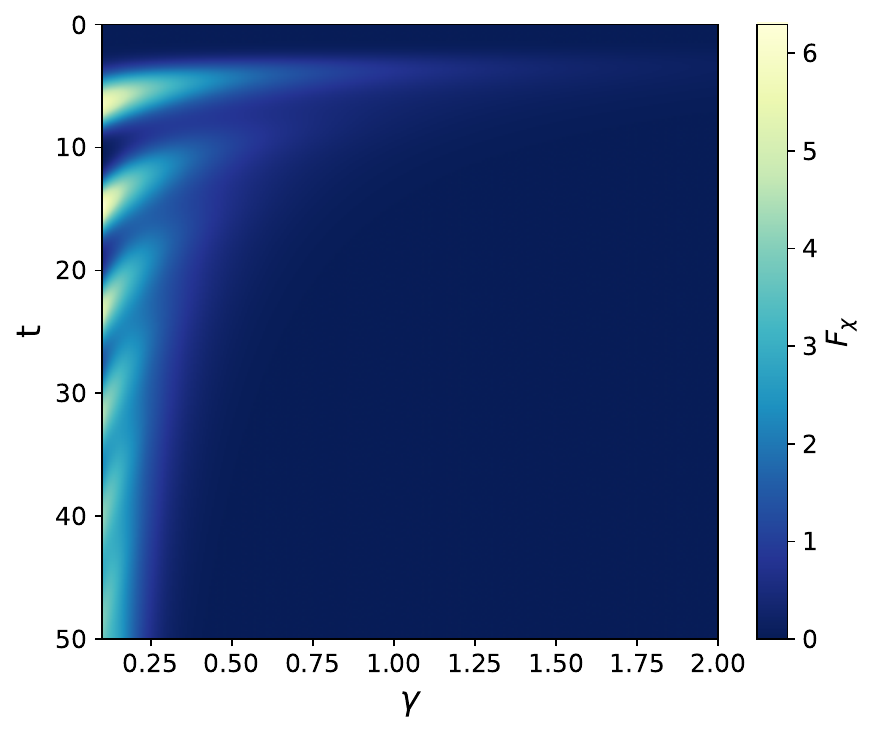}
    \captionof{figure}{Non-Markovian QFI for the initial state \(\ket{1,g}\) as a function
    of inverse bath memory time \(\gamma\) and evolution time \(t\); color
    represents \(F_\chi\). Although the closed and Markovian evolutions of this
    state yield zero QFI, finite environmental memory can activate sensitivity to
    \(\chi\). The scan exhibits appreciable sensitivity and temporal structure
    for smaller \(\gamma\), with suppression toward the short-memory side of the
    displayed range. No monotonic dependence on \(\gamma\) is implied.}
    \label{fig:memory_activated_qfi}
\end{minipage}
\end{center}

\onecolumngrid
\vfill
\begin{center}
\begin{minipage}{\textwidth}
    \centering
    \begin{minipage}[t]{0.49\textwidth}
        \centering
        \includegraphics[width=\linewidth]{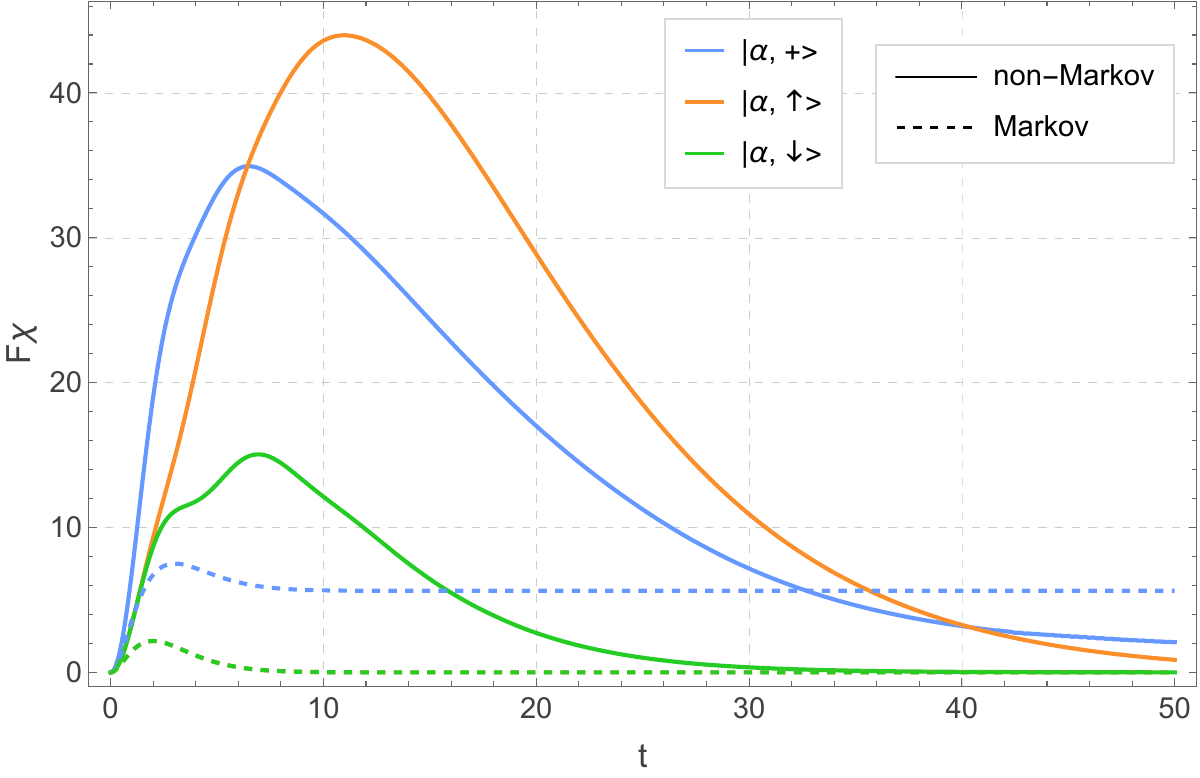}
        \vspace{0.25em}

        (a)
    \end{minipage}
    \hfill
    \begin{minipage}[t]{0.49\textwidth}
        \centering
        \includegraphics[width=\linewidth]{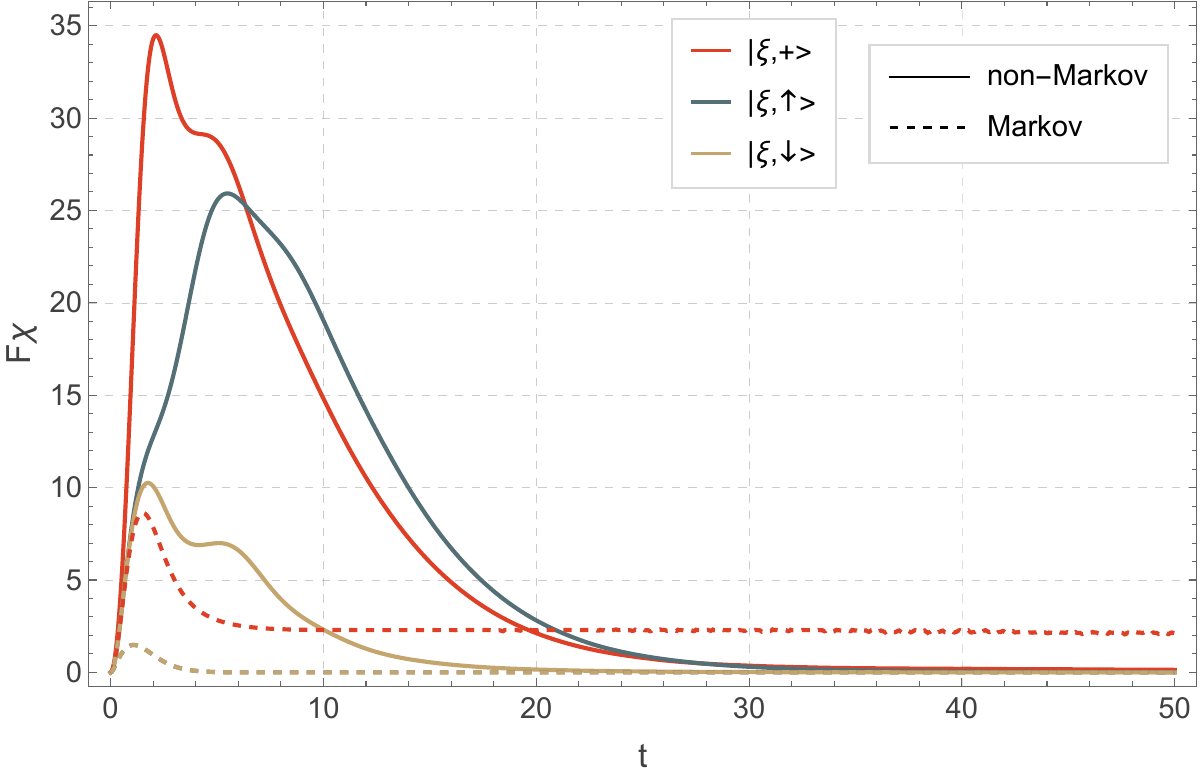}
        \vspace{0.25em}

        (b)
    \end{minipage}
    \captionof{figure}{QFI evolution for (a) coherent and (b) single-mode squeezed-vacuum
    cavity probes, each targeting mean photon number \(\langle\hat N\rangle=1\)
    and combined with the qubit states \(\ket{+}\), \(\ket e\), and \(\ket g\)
    (denoted by \(+\), \(\uparrow\), and \(\downarrow\) in the legends).
    The coherent state has \(\alpha=1\), and the squeezed vacuum has
    \(r=\operatorname{arcsinh}(1)\simeq0.881374\). Both calculations use a cavity
    dimension \(N_c=25\). Solid curves denote
    non-Markovian dynamics, for which \(\gamma/\Gamma=0.6\) and
    \(\Omega/\Gamma=0.2\), and dashed curves denote Markovian dynamics.
    In the Markovian regime, the curves for \(\ket{\alpha,\uparrow}\) and
    \(\ket{\alpha,\downarrow}\) coincide exactly in panel (a); likewise, the curves
    for \(\ket{\xi,\uparrow}\) and \(\ket{\xi,\downarrow}\) coincide exactly in
    panel (b).}
    \label{fig:v2-cohsqz}
\end{minipage}
\end{center}
\newpage
\twocolumngrid

To compare the role of photon-number statistics at fixed cavity excitation, we
consider coherent and single-mode squeezed-vacuum cavity states with the same mean
photon number, \(\langle\hat N\rangle=1\). We emphasize that the coherent state
\(\ket{\alpha=1}\) considered here is distinct from the single-photon Fock state
\(\ket{1}\) discussed above: the former is a superposition of Fock states with
Poissonian photon-number statistics, whereas the latter is a definite-number state
satisfying \(\hat N\ket{1}=\ket{1}\). For the coherent state,
\(\langle\hat N\rangle=|\alpha|^2\), and we therefore set \(\alpha=1\). For the
single-mode squeezed vacuum, \(\langle\hat N\rangle=\sinh^2 r\), so that
\(\langle\hat N\rangle=1\) corresponds to
\(r=\operatorname{arcsinh}(1)\simeq0.881374\).

Since both states occupy an infinite-dimensional Fock space, a finite cavity
truncation is required for the numerical calculations. Both the coherent state
\(\ket{\alpha=1}\) and the squeezed-vacuum state with \(r=0.881374\) are represented
using a cavity dimension \(N_c=25\), retaining
\(\{\ket{0},\ldots,\ket{24}\}\). The squeezed vacuum has support only on even
photon-number states, so only the even sectors of this truncation contribute.

Figure~\ref{fig:v2-cohsqz} shows the resulting QFI evolution for coherent probes
\(\ket{\alpha=1}\) and
squeezed-vacuum probes \(\ket\xi\), each combined with the qubit states \(\ket+\)
\(\ket e\), and \(\ket g\). The equal mean photon number removes the average cavity
occupation as the origin of their different performance. Their distinction instead
arises from their photon-number statistics. As Eq.~\eqref{eq:v2-closed-product-qfi}
shows, the closed-system QFI of a separable probe contains an explicit contribution from
\(\operatorname{Var}(\hat N)\), in addition to the contribution associated with qubit
coherence.

The numerical results show that these different photon-number distributions remain
important in the presence of environmental coupling. For coherent cavity probes, the
preferred qubit preparation depends more strongly on the bath dynamics:
\(\ket{\alpha,+}\) performs best among the coherent states in the Markovian regime,
while the relative ordering changes when finite environmental memory is introduced.
The squeezed probes display a more consistent hierarchy. In particular,
\(\ket{\xi,+}\) gives the largest QFI among the squeezed probes in both the Markovian
and non-Markovian regimes considered here. Hence, even at fixed
\(\langle\hat N\rangle\), the detailed photon-number statistics of the cavity field can
substantially alter the metrological response \cite{pezze2018quantum}.

Taken together, Figs.~\ref{fig:v2-qfi-states}--\ref{fig:v2-cohsqz} show that
closed-system generator variance remains a useful starting point for probe selection,
but is not a complete criterion in an open system. For most probes considered here,
useful sensitivity combines a strong initial response to \(G_\chi\) with coherence
and photon-number structure that remain robust under dissipation. The
\(\ket{1,+}\) probe combines these properties effectively in the single-photon
sector, while the coherent--squeezed comparison demonstrates the additional role of
photon-number statistics for more general cavity states.

The \(\ket{1,g}\) example reveals a further mechanism: finite environmental memory
can make the reduced dynamics themselves parameter-sensitive, activating finite QFI
for a probe that is inactive in the corresponding closed and Markovian dynamics.
Open-system probe selection therefore depends not only on the initial generator
structure, but also on how environmental dynamics encode, preserve, or return
information about \(\chi\). These complementary roles of probe preparation and
open-system evolution set the sensitivity benchmark for the measurement analysis
in the following section.

 ﻿\section{Optimal Measurement and Cram\'er--Rao Bound Achievability}
\label{sec:optimal_measurements}

The QFI determines the quantum Cram\'er--Rao bound for estimating the dispersive
coupling \(\chi\)~\cite{braunstein1994statistical},
\begin{equation}
(\Delta\chi)^2_{\mathrm{QCRB}}=\frac{1}{F_Q(\chi)}.
\label{eq:measurement_qcrb}
\end{equation}
For a Hermitian observable \(M\), the uncertainty obtained through error propagation is
\begin{equation}
(\Delta\chi)^2_M
=\frac{(\Delta M)^2}{\left|\partial_\chi\langle M\rangle\right|^2}
=\frac{\langle M^2\rangle-\langle M\rangle^2}
{\left|\partial_\chi\langle M\rangle\right|^2}.
\label{eq:measurement_error_propagation}
\end{equation}
The relation between observable fluctuations, parameter response, and quantum-limited
precision provides a basis for optimizing measurement observables through error
propagation~\cite{zhong2014optimal,frowis2015tighter}. We therefore minimize
\begin{equation}
\frac{\Delta M}{|\partial_\chi\langle M\rangle|}.
\end{equation}
An effective measurement should have sufficiently small fluctuations and a strong
response of its expectation value to changes in \(\chi\).

We consider the \(\ket{1,+}\) probe in the Markovian regime and numerically optimize
over a general Hermitian observable for near-optimal sensitivity throughout
\(0.50\leq\chi\leq0.60\). At the time corresponding to the maximum QFI, the resulting
optimized measurement operator is
\begin{align}
M_{\mathrm{opt}}={}&
-2.288\,I\otimes\sigma_x\nonumber\\
&-2.18643\,I\otimes\sigma_y
+0.719917\,\sigma_z\otimes I\nonumber\\
&+1.72027\,\sigma_z\otimes\sigma_x
+0.20526\,\sigma_z\otimes\sigma_y.
\label{eq:fixed_measurement_operator}
\end{align}
This expression provides an explicit measurement prescription in terms of
cavity--qubit observables. It is a single parameter-independent operator, optimized
over the chosen \(\chi\)-interval and subsequently kept fixed throughout that interval.
Adaptive estimation offers one route to adjusting measurements as information about
an unknown parameter accumulates~\cite{okamoto2012experimental}. Here, we instead
examine whether a single fixed observable maintains near-optimal local sensitivity
throughout a prescribed operating interval.

To quantify how closely the uncertainty from this fixed observable approaches the
quantum Cram\'er--Rao bound, we define the dimensionless ratio
\begin{equation}
R_{\mathrm{CRB}}(\chi)
=\frac{(\Delta\chi)^2_{\mathrm{QCRB}}}
{(\Delta\chi)^2_{M_{\mathrm{opt}}}}
=\frac{|\partial_\chi\langle M_{\mathrm{opt}}\rangle|^2}
{F_Q(\chi)(\Delta M_{\mathrm{opt}})^2}.
\label{eq:crb_achievability}
\end{equation}
The quantum Cram\'er--Rao bound implies \(0\leq R_{\mathrm{CRB}}\leq1\), with
\(R_{\mathrm{CRB}}=1\) corresponding to exact saturation. Values close to unity
indicate that the fixed observable gives an estimation uncertainty close to the
minimum permitted by the QFI.

\begin{figure}[t]
    \centering
    \includegraphics[width=\columnwidth]{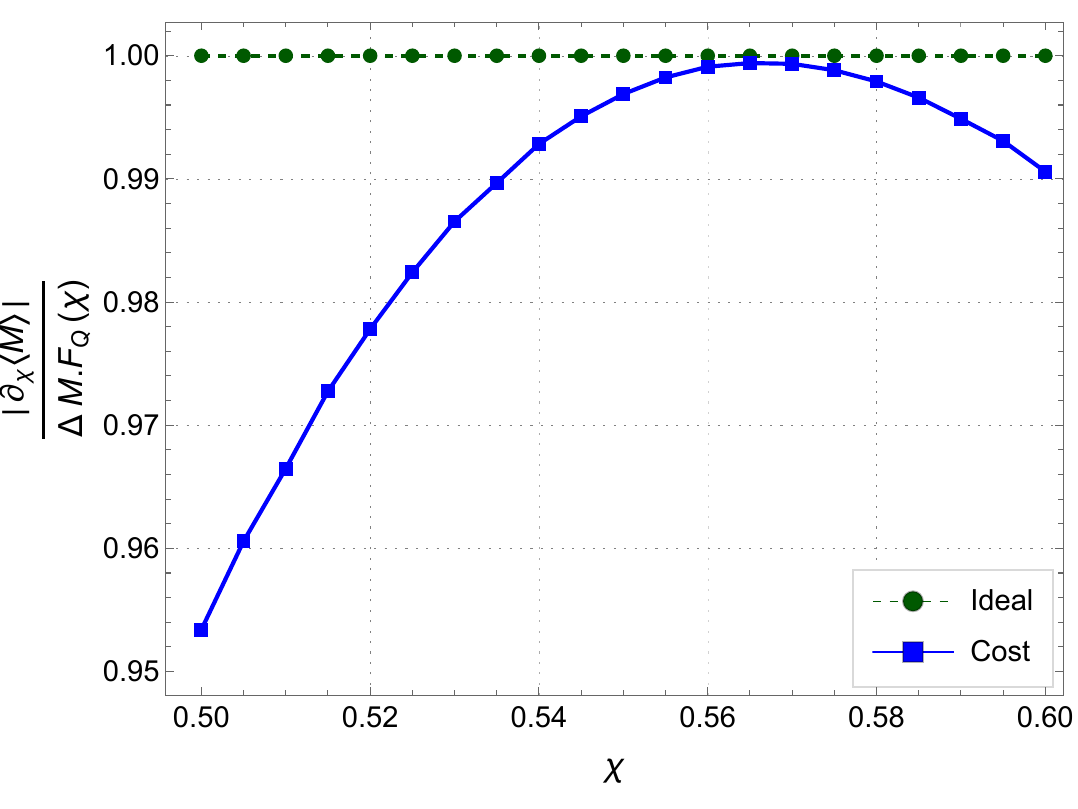}
    \caption{Achievability of the quantum Cram\'er--Rao bound for the
    \(\ket{1,+}\) probe in the Markovian regime. A general Hermitian measurement
    operator is optimized over \(0.50\leq\chi\leq0.60\) and subsequently held fixed.
    The ratio \(R_{\mathrm{CRB}}=(\Delta\chi)^2_{\mathrm{QCRB}}/
    (\Delta\chi)^2_{M_{\mathrm{opt}}}\) quantifies how closely the uncertainty
    obtained from this observable approaches the quantum Cram\'er--Rao bound.
    The optimized measurement gives \(R_{\mathrm{CRB}}>0.95\) throughout the
    interval. The dashed line at \(R_{\mathrm{CRB}}=1\) denotes exact saturation.}
    \label{fig:crb_achievability}
\end{figure}

Figure~\ref{fig:crb_achievability} shows \(R_{\mathrm{CRB}}(\chi)\) for the fixed
optimized Hermitian observable over \(0.50\leq\chi\leq0.60\). The observable is
optimized over this interval and then held fixed. The numerical result remains above
approximately \(R_{\mathrm{CRB}}\simeq0.95\) throughout the interval, indicating that
the fixed observable provides near-optimal local error-propagation sensitivity over
the full operating range without adjustment at each value of \(\chi\). This comparison
does not assert global optimality or guaranteed saturation by a finite-sample estimator.
 \section{Concluding Remarks and Discussion}

We have investigated the estimation of the dispersive coupling strength
\(\chi\) in an open cavity--qubit system, with particular emphasis on two
complementary aspects of the metrological protocol: the selection of the
initial probe and the measurement required to extract the encoded information.
In the closed-system limit, the generator \(G_\chi=\sigma_z\hat N\)
directly relates the attainable sensitivity to photon-number statistics and
qubit coherence. Although this generator structure provides a natural
benchmark for probe selection, the open-system results show that generator
variance alone is not sufficient: dissipative dynamics determine both the survival
of information-carrying components and, in finite-memory regimes, the encoding of
parameter dependence in the reduced state.

For the cavity-loss model considered here, the single-photon probe
\(\ket{1,+}\) retains the largest QFI among the representative states
considered in both Markovian and non-Markovian environments. The comparison
with the entangled family further demonstrates that entanglement alone is not
an indicator of metrological usefulness. A partially entangled state can
outperform the maximally entangled members of the same family, while the
separable \(\ket{1,+}\) probe can outperform the entire family considered
here. For these probes, the relevant quantity is the distribution of coherence among
eigenspaces of \(G_\chi\), together with the robustness of that coherence
against photon loss. The coherent--squeezed comparison reinforces this
conclusion from a complementary perspective: even at fixed mean photon
number, different photon-number statistics lead to substantially different
sensitivities. In particular, \(\ket{\xi,+}\) yields the largest QFI among
the squeezed probes in both the Markovian and non-Markovian regimes considered
here.

Environmental memory provides an additional means of preserving metrological
information. In the Markovian regime, photon loss strongly suppresses the
attainable QFI, whereas finite bath memory can substantially increase both the
magnitude and lifetime of the sensitivity. The exact non-Markovian
\(O\)-operator construction allows this enhancement to be associated
directly with the finite correlation time of the environment, while
recovering the Markovian dynamics in the short-memory limit. The
\(\ket{1,g}\) example also demonstrates memory-induced activation of sensitivity:
although this probe has zero closed-system and Markovian QFI, the
\(\chi\)-dependent bath response can produce distinguishable reduced populations
for suitable finite-memory parameters. Environmental memory therefore modifies both
the retention and the dynamical encoding of parameter information, without changing
the initial probe's closed-system generator variance.

The stochastic and density-operator descriptions provide complementary
perspectives on this open-system sensitivity. The SQFI quantifies the
information retained in individual conditioned trajectories before the
environmental record is discarded, whereas the density-operator QFI
characterizes the information available from the unconditional reduced state.
Their comparison therefore distinguishes trajectory-resolved information from
the information surviving noise averaging. The slightly larger long-time SQFI
found in the Markovian calculations is consistent with this distinction and
with the convexity of the QFI under parameter-independent mixing.

We have further connected these sensitivity bounds to an explicit measurement
prescription by numerically optimizing a general Hermitian observable. For the
\(\ket{1,+}\) probe in the Markovian regime, a single fixed measurement is used
throughout \(0.50\leq\chi\leq0.60\). The fixed optimized observable yields an
error-propagation sensitivity ratio \(R_{\mathrm{CRB}}>0.95\) throughout the parameter
interval considered, indicating near-optimal sensitivity without readjusting the
observable at each coupling value.

Future work will extend this framework to the simultaneous estimation of the
dispersive coupling and environmental parameters, such as the bath correlation rate
and photon-loss strength. This will require examining parameter correlations through
the quantum Fisher information matrix, assessing measurement compatibility, and
determining attainable precision bounds, including the Holevo bound when
appropriate~\cite{liu2020quantum,ragy2016compatibility,albarelli2019holevo}. In
particular, the inverse quantum Fisher information matrix is not automatically a
simultaneously attainable precision bound when the optimal measurements for different
parameters are incompatible.

A complementary direction is to develop efficient methods for evaluating and
experimentally certifying the metrological information in the evolving cavity--qubit
state. Alternative matrix formulations provide a starting point for numerical
evaluation without explicit density-matrix diagonalization~\cite{safranek2018simple},
but do not by themselves guarantee a universal computational speedup. Randomized-
measurement protocols can provide access to the QFI in suitable settings or to
systematically improvable lower bounds without full state tomography
\cite{rath2021randomized,yu2021experimental}. Efficiently estimable surrogates such as
sub-QFI also warrant investigation for probe optimization and sensitivity
certification~\cite{cerezo2021sub}. Such lower bounds and surrogates must remain
distinguished from the exact QFI, and surrogate-optimality results established for
unitary parameter families cannot be assumed to carry over to the dissipative
dynamics considered here.

Taken together, these results suggest a simple design principle for dispersive
open-system metrology: probe selection should account for the initial generator
response, the robustness of coherence and photon-number structure, and any
parameter dependence encoded by the environmental dynamics, while the associated
measurement should remain sufficiently stable over the expected operating
range. Environmental memory can preserve or return information lost through
Markovian decay and, for suitable bath parameters, activate sensitivity in an
otherwise insensitive probe.
These considerations provide a framework for cavity- and circuit-QED
implementations, where dispersive interactions, qubit rotations, cavity
readout, and reservoir engineering are experimentally accessible. Natural
extensions include finite-temperature reservoirs, detector inefficiency,
and adaptive measurement strategies over wider parameter ranges.
 \begin{acknowledgments}
We thank Prof. Yogesh Joglekar for his insightful comments and suggestions, and
Dr. Kenneth Mui and Duan Wang for helpful discussions. This project is supported
by ACC-New Jersey under Contract No. W15QKN-24-C-0004.
\end{acknowledgments}
 \appendix

\section{Markovian QSD Derivation for Stochastic QFI}
\label{app:v2-markovian-qsd-sqfi}

For the density-operator benchmark in the Markovian regime, the QFI is evaluated from
the spectral expression
\begin{equation}
\mathcal F_\chi
=2\sum_{m,n}
\frac{\left|\bra{m}\partial_\chi\rho_\chi\ket{n}\right|^2}{\lambda_m+\lambda_n},
\label{eq:v2-app-density-qfi}
\end{equation}
where \(\rho_\chi=\sum_m \lambda_m\ket{m}\bra{m}\), and terms with
\(\lambda_m+\lambda_n=0\) are omitted.
The stochastic calculation uses the quantum-state-diffusion unraveling of the same
Lindblad dynamics \cite{gisin1992quantum,percival1998quantum}. In units where the
Markovian decay rate is scaled to one, the trajectory state obeys
\begin{equation}
    \partial_t\ket{\psi_t}
    =
    \left[-iH_S-\frac{1}{2}a^\dagger a+a z_t^*\right]\ket{\psi_t},
    \label{eq:v2-app-markovian-qsd-main}
\end{equation}
with complex Gaussian white noise
\begin{equation}
    \mathcal{M}[z_t^*z_s]=\delta(t-s),\qquad
    \mathcal{M}[z_tz_s]=0,\qquad
    \mathcal{M}[z_t]=0.
\end{equation}
The stochastic QFI for a single pure trajectory is
\begin{equation}
\mathcal F_\chi[z^*]
=4\left(\braket{\partial_\chi\psi|\partial_\chi\psi}
-|\braket{\psi|\partial_\chi\psi}|^2\right).
\label{eq:v2-app-stochastic-qfi-main}
\end{equation}
Since the noise and the parameter \(\chi\) are independent, the derivative
\(\ket{\phi_t}=\partial_\chi\ket{\psi_t}\) obeys
\begin{equation}
    \partial_t\ket{\phi_t}
    =
    \left[-iH_S-\frac{1}{2}a^\dagger a+a z_t^*\right]\ket{\phi_t}
    -i\left(\partial_\chi H_S\right)\ket{\psi_t}.
    \label{eq:v2-app-phi-evolution}
\end{equation}
The physical SQFI estimate is obtained by evaluating
Eq.~\eqref{eq:v2-app-stochastic-qfi-main} on each trajectory and averaging over the
noise realizations.

Expanding the trajectory in the joint photon-qubit basis,
\begin{equation}
    \ket{\psi_t}=\sum_{n=0}^{\infty}\sum_{q=\pm 1}
    c_{n,q}(t,\chi,z^*)\ket{n,q},
\end{equation}
one obtains the recursive coefficient equation
\begin{align}
    \dot c_{n,q}
    &=
    \left[-\frac{n}{2}-in\omega_c-iq(\omega_q/2+\chi n)\right]c_{n,q}
    \nonumber\\
    &\quad
    +z_t^*\sqrt{n+1}\,c_{n+1,q}.
    \label{eq:v2-app-coefficient-recursion}
\end{align}
For the single-photon truncation, \(c_{2,q}=0\). For the normalized balanced initial
state
\(\ket{\psi_0}=(\ket{0,g}+\ket{0,e}+\ket{1,g}+\ket{1,e})/2\), the coefficients are
\begin{align}
    c_{1,q}(t)
    &=
    \frac{1}{2}
    e^{-t/2}e^{-i[\omega_c+q(\omega_q/2+\chi)]t},
    \\
    c_{0,q}(t)
    &=
    \frac{1}{2}e^{-iq\omega_q t/2}
    \left[
    1+\int_0^t ds\,
    z_s^* e^{-s/2}e^{-i(\omega_c+q\chi)s}
    \right].
    \label{eq:v2-app-single-photon-coefficients}
\end{align}
Differentiating with respect to \(\chi\) gives
\begin{align}
    \partial_\chi c_{1,q}(t)
    &=
    -iqt\,c_{1,q}(t),
    \\
    \partial_\chi c_{0,q}(t)
    &=
    -\frac{iq}{2}e^{-iq\omega_q t/2}
    \int_0^t ds\,
    s z_s^* e^{-s/2}e^{-i(\omega_c+q\chi)s}.
\end{align}
Using orthogonality of the basis states, the first contribution to the SQFI becomes
\begin{align}
    \braket{\phi_t|\phi_t}
    &=
    \frac{1}{4}
    \int_0^t ds\int_0^t ds'\,
    z_s z_{s'}^*ss'e^{-(s+s')/2}
    \nonumber\\
    &\quad\times
    \left(
    e^{i(\omega_c+\chi)(s-s')}
    +
    e^{i(\omega_c-\chi)(s-s')}
    \right)
    +\frac{1}{2}t^2e^{-t}.
\end{align}
Taking the ensemble average gives
\begin{equation}
    \mathcal{M}\!\left[\braket{\phi_t|\phi_t}\right]
    =
    1-e^{-t}(1+t).
    \label{eq:v2-app-phi-phi-average}
\end{equation}

The second contribution can be written compactly by defining
\begin{align}
    f(s)&=2s e^{-s/2}e^{-i\omega_c s}\sin(\chi s),\\
    g(s,s')&=2s'e^{-(s+s')/2}e^{i\omega_c(s-s')}
    \sin\!\left[\chi(s-s')\right].
\end{align}
For the normalized initial state,
\begin{align}
    \braket{\psi_t|\phi_t}
    &=
    -\frac{1}{4}\int_0^t ds\,z_s^* f(s)
    \nonumber\\
    &\quad
    +\frac{1}{4}\int_0^t ds\int_0^t ds'\,
    z_s z_{s'}^*g(s,s').
\end{align}
Odd moments of the Gaussian noise vanish, while the four-point function factorizes as
\begin{equation}
    \mathcal{M}[z_a z_b^* z_c z_d^*]
    =
    \mathcal{M}[z_a z_b^*]\mathcal{M}[z_c z_d^*]
    +
    \mathcal{M}[z_a z_d^*]\mathcal{M}[z_c z_b^*].
\end{equation}
Since \(g(s,s)=0\), this yields
\begin{align}
    \mathcal{M}\!\left[\left|\braket{\psi_t|\phi_t}\right|^2\right]
    &=
    \frac{1}{16}
    \left[
    \int_0^t ds\,|f(s)|^2
    \right.
    \nonumber\\
    &\quad\left.
    +
    \int_0^t ds\int_0^t ds'\,|g(s,s')|^2
    \right],
    \label{eq:v2-app-overlap-average}
\end{align}
where
\begin{align}
    |f(s)|^2 &= 4s^2 e^{-s}\sin^2(\chi s),\\
    |g(s,s')|^2 &=4{s'}^2 e^{-(s+s')}
    \sin^2\!\left[\chi(s-s')\right].
\end{align}
Combining Eqs.~\eqref{eq:v2-app-stochastic-qfi-main},
\eqref{eq:v2-app-phi-phi-average}, and \eqref{eq:v2-app-overlap-average}, the
ensemble-averaged SQFI in the single-photon truncation is
\begin{equation}
\begin{split}
    \overline{\mathcal{F}}_\chi(t)
    &=
    4\left[1-e^{-t}(1+t)\right]
    -\int_0^t ds\,s^2e^{-s}\sin^2(\chi s)
    \\
    &\quad
    -\int_0^t ds\int_0^t ds'\,
    {s'}^2e^{-(s+s')}\sin^2\!\left[\chi(s-s')\right].
\end{split}
\label{eq:v2-app-markovian-sqfi-final}
\end{equation}
This analytical trajectory average is the benchmark used to validate the numerical
average over a finite ensemble of QSD trajectories. It also makes explicit why stochastic QFI
can carry trajectory-level conditional information before the ensemble average is
formed.

\section{Non-Markovian O-Operator Solution}
\label{app:v2-nm-o-operator}

The non-Markovian calculation uses the exact quantum-state-diffusion construction
\cite{diosi1998non,strunz1999open,yu1999non}. The stochastic state satisfies

\begin{figure}[t]
    \centering
    \includegraphics[width=0.96\columnwidth]{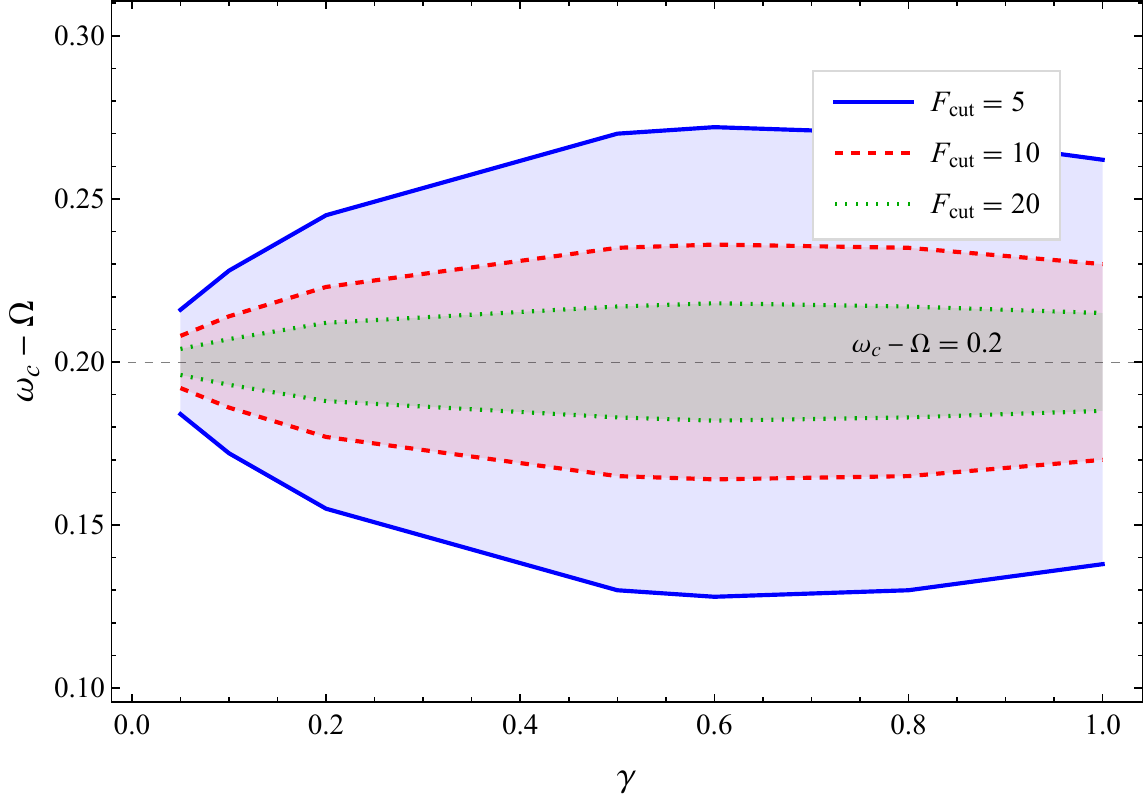}
    \caption{Numerically determined excluded region in the
    \((\gamma,\omega_c-\Omega)\) plane associated with the \(X_-(t)\) branch.
    The lower and upper boundaries correspond to the condition
    \(\max_t|X_-(t)|=F_{\mathrm{cut}}\), with the cutoff values shown in the
    figure. The shaded regions indicate parameter values for which the maximum
    magnitude of \(X_-(t)\) exceeds the corresponding cutoff. The dashed
    horizontal line marks \((\omega_c-\Omega)/\Gamma=\chi/\Gamma=0.2\), around
    which the pole-like enhancement of the \(X_-\) branch is localized. The
    corresponding \(X_+\) enhancement occurs at
    \((\omega_c-\Omega)/\Gamma=-\chi/\Gamma\).}
    \label{fig:v2-app-nm-parameter-bounds}
\end{figure}

\begin{equation}
    \frac{d}{dt}\ket{\psi_t}
    =
    \left[
    -iH_S+\sqrt{\Gamma}\,a z_t^*
    -\sqrt{\Gamma} a^\dagger\overline O(t)
    \right]\ket{\psi_t},
    \label{eq:v2-app-nmqsd}
\end{equation}
with
\begin{equation}
    \mathcal{M}[z_t^*z_s]=\alpha(t,s),\qquad
    \mathcal{M}[z_tz_s]=0,\qquad
    \mathcal{M}[z_t]=0.
\end{equation}
Introducing
\begin{equation}
    \frac{\delta\ket{\psi_t}}{\delta z_s^*}
    = O(t,s)\ket{\psi_t},
\end{equation}
the consistency condition gives
\begin{equation}
    \dot{O}
    =
    \left[-iH_S+\sqrt{\Gamma}\,a z_t^*
    -\sqrt{\Gamma} a^\dagger\overline O,\,O\right]
    -\sqrt{\Gamma}\,a^\dagger
    \frac{\delta \overline O}{\delta z_s^*},
    \label{eq:v2-app-nm-o-consistency}
\end{equation}
where
\begin{equation}
    \overline O(t)=\int_0^t ds\,\alpha(t,s)O(t,s).
    \label{eq:v2-app-nm-bar-o}
\end{equation}
The corresponding time-local master equation is
\begin{equation}
    \dot{\rho}=-i[H_S,\rho]
    +\sqrt{\Gamma}[a,\rho\overline O^\dagger(t)]
    +\sqrt{\Gamma}[\overline O(t)\rho,a^\dagger].
    \label{eq:v2-app-nm-master}
\end{equation}

For the dispersive model, we use the noise-independent ansatz
\begin{equation}
    O(t,s)=f_1(t,s)a+f_2(t,s)a\sigma_z ,
    \label{eq:v2-app-nm-o-ansatz}
\end{equation}
with boundary conditions \(f_1(t,t)=\sqrt{\Gamma}\) and \(f_2(t,t)=0\). This ansatz is closed
because the commutator with the system Hamiltonian produces only the same two operator
structures,
\begin{align}
    [H_S,O]
    &=
    -(\omega_c f_1+\chi f_2)a
    \nonumber\\
    &\quad
    -(\omega_c f_2+\chi f_1)a\sigma_z,
\end{align}
and \([a,O]=0\). Defining
\begin{equation}
    F_i(t)=\int_0^t ds\,\alpha(t,s)f_i(t,s),\qquad i=1,2,
    \label{eq:v2-app-Fi-definition}
\end{equation}
one has
\begin{equation}
    \overline O(t)=F_1(t)a+F_2(t)a\sigma_z .
\end{equation}
Equating the coefficients of \(a\) and \(a\sigma_z\) in
Eq.~\eqref{eq:v2-app-nm-o-consistency} gives
\begin{align}
    \dot f_1
    &=
    i(\omega_cf_1+\chi f_2)
    +\sqrt{\Gamma}(F_1f_1+F_2f_2),
    \label{eq:v2-app-f1dot}\\
    \dot f_2
    &=
    i(\omega_cf_2+\chi f_1)
    +\sqrt{\Gamma}(F_1f_2+F_2f_1).
    \label{eq:v2-app-f2dot}
\end{align}

For the Ornstein--Uhlenbeck kernel
\begin{equation}
    \alpha(t,s)=\frac{\gamma}{2}e^{-\gamma|t-s|-i\Omega(t-s)},
\end{equation}
we have \(\alpha(t,t)=\gamma/2 \) and
\(\partial_t\alpha(t,s)=(-\gamma-i\Omega)\alpha(t,s)\) for \(t>s\). The combinations
\(X_+=F_1+F_2\) and \(X_-=F_1-F_2\) obey decoupled Riccati equations,
\begin{align}
    \dot X_+&=\frac{\gamma}{2}+\xi_+X_+ +\sqrt{\Gamma} X_+^2,\\
    \dot X_-&=\frac{\gamma}{2}+\xi_-X_- +\sqrt{\Gamma} X_-^2,
\end{align}
where
\begin{equation}
    \xi_\pm=-\gamma+i(\omega_c-\Omega\pm\chi).
\end{equation}
Completing the square gives
\begin{equation}
    \dot X_\pm
    =
    \sqrt{\Gamma}\left[
    \left(X_\pm+\frac{\xi_\pm}{2\sqrt{\Gamma}}\right)^2-\delta_\pm^2
    \right],
\end{equation}
with
\begin{equation}
    \delta_\pm^2=\frac{\xi_\pm^2}{4\Gamma}-\frac{\gamma}{2}.
\end{equation}
The solutions are
\begin{equation}
    X_\pm(t)
    =
    \delta_\pm\left(\frac{2}{1-c_\pm e^{2\sqrt{\Gamma}\delta_\pm t}}-1\right)
    -\frac{\xi_\pm}{2\sqrt{\Gamma}},
    \label{eq:v2-app-Xpm-solution}
\end{equation}
where
\begin{equation}
    c_\pm=
    \frac{\frac{\xi_\pm}{2\sqrt{\Gamma}}-\delta_\pm}
    {\frac{\xi_\pm}{2\sqrt{\Gamma}}+\delta_\pm}.
\end{equation}
Finally,
\begin{equation}
    F_1(t)=\frac{X_+(t)+X_-(t)}{2},\qquad
    F_2(t)=\frac{X_+(t)-X_-(t)}{2}.
\end{equation}
Equivalently, the memory coefficients can be written explicitly as
\begin{widetext}
\begin{align}
    F_1(t)
    &=
    \frac{\delta_+}{2}\left(\frac{2}{1-c_+e^{2\sqrt{\Gamma}\delta_+ t}}-1\right)
    +
    \frac{\delta_-}{2}\left(\frac{2}{1-c_-e^{2\sqrt{\Gamma}\delta_- t}}-1\right)
    -\frac{\xi_++\xi_-}{4\sqrt{\Gamma}},
    \label{eq:v2-app-nm-F1-exact}\\
    F_2(t)
    &=
    \frac{\delta_+}{2}\left(\frac{2}{1-c_+e^{2\sqrt{\Gamma}\delta_+ t}}-1\right)
    -
    \frac{\delta_-}{2}\left(\frac{2}{1-c_-e^{2\sqrt{\Gamma}\delta_- t}}-1\right)
    -\frac{\xi_+-\xi_-}{4\sqrt{\Gamma}}.
    \label{eq:v2-app-nm-F2-exact}
\end{align}
\end{widetext}
These coefficients determine the non-Markovian decay and revival of the coherences that
carry information about \(\chi\).

Figure~\ref{fig:v2-app-nm-parameter-bounds} summarizes the numerical stability bounds
associated with the pole-like enhancement of the \(X_-\) branch around
\(\omega_c-\Omega=\chi\), over the range of bath memory parameters considered in
the main text. The analogous \(X_+\) enhancement is centered at
\(\omega_c-\Omega=-\chi\).


\begin{thebibliography}{50}%
\makeatletter
\providecommand \@ifxundefined [1]{%
 \@ifx{#1\undefined}
}%
\providecommand \@ifnum [1]{%
 \ifnum #1\expandafter \@firstoftwo
 \else \expandafter \@secondoftwo
 \fi
}%
\providecommand \@ifx [1]{%
 \ifx #1\expandafter \@firstoftwo
 \else \expandafter \@secondoftwo
 \fi
}%
\providecommand \natexlab [1]{#1}%
\providecommand \enquote  [1]{``#1''}%
\providecommand \bibnamefont  [1]{#1}%
\providecommand \bibfnamefont [1]{#1}%
\providecommand \citenamefont [1]{#1}%
\providecommand \href@noop [0]{\@secondoftwo}%
\providecommand \href [0]{\begingroup \@sanitize@url \@href}%
\providecommand \@href[1]{\@@startlink{#1}\@@href}%
\providecommand \@@href[1]{\endgroup#1\@@endlink}%
\providecommand \@sanitize@url [0]{\catcode `\\12\catcode `\$12\catcode
  `\&12\catcode `\#12\catcode `\^12\catcode `\_12\catcode `\%12\relax}%
\providecommand \@@startlink[1]{}%
\providecommand \@@endlink[0]{}%
\providecommand \url  [0]{\begingroup\@sanitize@url \@url }%
\providecommand \@url [1]{\endgroup\@href {#1}{\urlprefix }}%
\providecommand \urlprefix  [0]{URL }%
\providecommand \Eprint [0]{\href }%
\providecommand \doibase [0]{https://doi.org/}%
\providecommand \selectlanguage [0]{\@gobble}%
\providecommand \bibinfo  [0]{\@secondoftwo}%
\providecommand \bibfield  [0]{\@secondoftwo}%
\providecommand \translation [1]{[#1]}%
\providecommand \BibitemOpen [0]{}%
\providecommand \bibitemStop [0]{}%
\providecommand \bibitemNoStop [0]{.\EOS\space}%
\providecommand \EOS [0]{\spacefactor3000\relax}%
\providecommand \BibitemShut  [1]{\csname bibitem#1\endcsname}%
\let\auto@bib@innerbib\@empty
\bibitem [{\citenamefont {Rao}\ \emph {et~al.}(1945)\citenamefont {Rao} \emph
  {et~al.}}]{rao1945information}%
  \BibitemOpen
  \bibfield  {author} {\bibinfo {author} {\bibfnamefont {C.~R.}\ \bibnamefont
  {Rao}} \emph {et~al.},\ }\bibfield  {title} {\bibinfo {title} {Information
  and the accuracy attainable in the estimation of statistical parameters},\
  }\href@noop {} {\bibfield  {journal} {\bibinfo  {journal} {Bull. Calcutta
  Math. Soc}\ }\textbf {\bibinfo {volume} {37}},\ \bibinfo {pages} {81}
  (\bibinfo {year} {1945})}\BibitemShut {NoStop}%
\bibitem [{\citenamefont {Helstrom}(1967)}]{helstrom1967detection}%
  \BibitemOpen
  \bibfield  {author} {\bibinfo {author} {\bibfnamefont {C.~W.}\ \bibnamefont
  {Helstrom}},\ }\bibfield  {title} {\bibinfo {title} {Detection theory and
  quantum mechanics},\ }\href@noop {} {\bibfield  {journal} {\bibinfo
  {journal} {Information and Control}\ }\textbf {\bibinfo {volume} {10}},\
  \bibinfo {pages} {254} (\bibinfo {year} {1967})}\BibitemShut {NoStop}%
\bibitem [{\citenamefont {Helstrom}(1969)}]{helstrom1969quantum}%
  \BibitemOpen
  \bibfield  {author} {\bibinfo {author} {\bibfnamefont {C.~W.}\ \bibnamefont
  {Helstrom}},\ }\bibfield  {title} {\bibinfo {title} {Quantum detection and
  estimation theory},\ }\href@noop {} {\bibfield  {journal} {\bibinfo
  {journal} {Journal of statistical physics}\ }\textbf {\bibinfo {volume}
  {1}},\ \bibinfo {pages} {231} (\bibinfo {year} {1969})}\BibitemShut {NoStop}%
\bibitem [{\citenamefont {Braunstein}\ and\ \citenamefont
  {Caves}(1994)}]{braunstein1994statistical}%
  \BibitemOpen
  \bibfield  {author} {\bibinfo {author} {\bibfnamefont {S.~L.}\ \bibnamefont
  {Braunstein}}\ and\ \bibinfo {author} {\bibfnamefont {C.~M.}\ \bibnamefont
  {Caves}},\ }\bibfield  {title} {\bibinfo {title} {Statistical distance and
  the geometry of quantum states},\ }\href
  {https://doi.org/10.1103/PhysRevLett.72.3439} {\bibfield  {journal} {\bibinfo
   {journal} {Physical Review Letters}\ }\textbf {\bibinfo {volume} {72}},\
  \bibinfo {pages} {3439} (\bibinfo {year} {1994})}\BibitemShut {NoStop}%
\bibitem [{\citenamefont {Paris}(2009)}]{paris2009quantum}%
  \BibitemOpen
  \bibfield  {author} {\bibinfo {author} {\bibfnamefont {M.~G.~A.}\
  \bibnamefont {Paris}},\ }\bibfield  {title} {\bibinfo {title} {Quantum
  estimation for quantum technology},\ }\href
  {https://doi.org/10.1142/S0219749909004839} {\bibfield  {journal} {\bibinfo
  {journal} {International Journal of Quantum Information}\ }\textbf {\bibinfo
  {volume} {7}},\ \bibinfo {pages} {125} (\bibinfo {year} {2009})}\BibitemShut
  {NoStop}%
\bibitem [{\citenamefont {Giovannetti}\ \emph {et~al.}(2006)\citenamefont
  {Giovannetti}, \citenamefont {Lloyd},\ and\ \citenamefont
  {Maccone}}]{giovannetti2006quantum}%
  \BibitemOpen
  \bibfield  {author} {\bibinfo {author} {\bibfnamefont {V.}~\bibnamefont
  {Giovannetti}}, \bibinfo {author} {\bibfnamefont {S.}~\bibnamefont {Lloyd}},\
  and\ \bibinfo {author} {\bibfnamefont {L.}~\bibnamefont {Maccone}},\
  }\bibfield  {title} {\bibinfo {title} {Quantum metrology},\ }\href@noop {}
  {\bibfield  {journal} {\bibinfo  {journal} {Physical review letters}\
  }\textbf {\bibinfo {volume} {96}},\ \bibinfo {pages} {010401} (\bibinfo
  {year} {2006})}\BibitemShut {NoStop}%
\bibitem [{\citenamefont {Giovannetti}\ \emph {et~al.}(2011)\citenamefont
  {Giovannetti}, \citenamefont {Lloyd},\ and\ \citenamefont
  {Maccone}}]{giovannetti2011advances}%
  \BibitemOpen
  \bibfield  {author} {\bibinfo {author} {\bibfnamefont {V.}~\bibnamefont
  {Giovannetti}}, \bibinfo {author} {\bibfnamefont {S.}~\bibnamefont {Lloyd}},\
  and\ \bibinfo {author} {\bibfnamefont {L.}~\bibnamefont {Maccone}},\
  }\bibfield  {title} {\bibinfo {title} {Advances in quantum metrology},\
  }\href@noop {} {\bibfield  {journal} {\bibinfo  {journal} {Nature photonics}\
  }\textbf {\bibinfo {volume} {5}},\ \bibinfo {pages} {222} (\bibinfo {year}
  {2011})}\BibitemShut {NoStop}%
\bibitem [{\citenamefont {T{\'o}th}\ and\ \citenamefont
  {Apellaniz}(2014)}]{toth2014quantum}%
  \BibitemOpen
  \bibfield  {author} {\bibinfo {author} {\bibfnamefont {G.}~\bibnamefont
  {T{\'o}th}}\ and\ \bibinfo {author} {\bibfnamefont {I.}~\bibnamefont
  {Apellaniz}},\ }\bibfield  {title} {\bibinfo {title} {Quantum metrology from
  a quantum information science perspective},\ }\href@noop {} {\bibfield
  {journal} {\bibinfo  {journal} {Journal of Physics A: Mathematical and
  Theoretical}\ }\textbf {\bibinfo {volume} {47}},\ \bibinfo {pages} {424006}
  (\bibinfo {year} {2014})}\BibitemShut {NoStop}%
\bibitem [{\citenamefont {Pang}\ and\ \citenamefont
  {Brun}(2014)}]{pang2014quantum}%
  \BibitemOpen
  \bibfield  {author} {\bibinfo {author} {\bibfnamefont {S.}~\bibnamefont
  {Pang}}\ and\ \bibinfo {author} {\bibfnamefont {T.~A.}\ \bibnamefont
  {Brun}},\ }\bibfield  {title} {\bibinfo {title} {Quantum metrology for a
  general hamiltonian parameter},\ }\href
  {https://doi.org/10.1103/PhysRevA.90.022117} {\bibfield  {journal} {\bibinfo
  {journal} {Physical Review A}\ }\textbf {\bibinfo {volume} {90}},\ \bibinfo
  {pages} {022117} (\bibinfo {year} {2014})}\BibitemShut {NoStop}%
\bibitem [{\citenamefont {Escher}\ \emph {et~al.}(2011)\citenamefont {Escher},
  \citenamefont {de~Matos~Filho},\ and\ \citenamefont
  {Davidovich}}]{escher2011general}%
  \BibitemOpen
  \bibfield  {author} {\bibinfo {author} {\bibfnamefont {B.~M.}\ \bibnamefont
  {Escher}}, \bibinfo {author} {\bibfnamefont {R.~L.}\ \bibnamefont
  {de~Matos~Filho}},\ and\ \bibinfo {author} {\bibfnamefont {L.}~\bibnamefont
  {Davidovich}},\ }\bibfield  {title} {\bibinfo {title} {General framework for
  estimating the ultimate precision limit in noisy quantum-enhanced
  metrology},\ }\href {https://doi.org/10.1038/nphys1958} {\bibfield  {journal}
  {\bibinfo  {journal} {Nature Physics}\ }\textbf {\bibinfo {volume} {7}},\
  \bibinfo {pages} {406} (\bibinfo {year} {2011})}\BibitemShut {NoStop}%
\bibitem [{\citenamefont {Demkowicz-Dobrza{\'n}ski}\ \emph
  {et~al.}(2012)\citenamefont {Demkowicz-Dobrza{\'n}ski}, \citenamefont
  {Ko{\l}ody{\'n}ski},\ and\ \citenamefont
  {Gu{\c{t}}{\u{a}}}}]{demkowicz2012elusive}%
  \BibitemOpen
  \bibfield  {author} {\bibinfo {author} {\bibfnamefont {R.}~\bibnamefont
  {Demkowicz-Dobrza{\'n}ski}}, \bibinfo {author} {\bibfnamefont
  {J.}~\bibnamefont {Ko{\l}ody{\'n}ski}},\ and\ \bibinfo {author}
  {\bibfnamefont {M.}~\bibnamefont {Gu{\c{t}}{\u{a}}}},\ }\bibfield  {title}
  {\bibinfo {title} {The elusive {H}eisenberg limit in quantum-enhanced
  metrology},\ }\href {https://doi.org/10.1038/ncomms2067} {\bibfield
  {journal} {\bibinfo  {journal} {Nature Communications}\ }\textbf {\bibinfo
  {volume} {3}},\ \bibinfo {pages} {1063} (\bibinfo {year} {2012})}\BibitemShut
  {NoStop}%
\bibitem [{\citenamefont {Lindblad}(1976)}]{lindblad1976generators}%
  \BibitemOpen
  \bibfield  {author} {\bibinfo {author} {\bibfnamefont {G.}~\bibnamefont
  {Lindblad}},\ }\bibfield  {title} {\bibinfo {title} {On the generators of
  quantum dynamical semigroups},\ }\href@noop {} {\bibfield  {journal}
  {\bibinfo  {journal} {Communications in mathematical physics}\ }\textbf
  {\bibinfo {volume} {48}},\ \bibinfo {pages} {119} (\bibinfo {year}
  {1976})}\BibitemShut {NoStop}%
\bibitem [{\citenamefont {Breuer}\ \emph {et~al.}(2009)\citenamefont {Breuer},
  \citenamefont {Laine},\ and\ \citenamefont {Piilo}}]{breuer2009measure}%
  \BibitemOpen
  \bibfield  {author} {\bibinfo {author} {\bibfnamefont {H.-P.}\ \bibnamefont
  {Breuer}}, \bibinfo {author} {\bibfnamefont {E.-M.}\ \bibnamefont {Laine}},\
  and\ \bibinfo {author} {\bibfnamefont {J.}~\bibnamefont {Piilo}},\ }\bibfield
   {title} {\bibinfo {title} {Measure for the degree of non-markovian behavior
  of quantum processes in open systems},\ }\href
  {https://doi.org/10.1103/PhysRevLett.103.210401} {\bibfield  {journal}
  {\bibinfo  {journal} {Physical Review Letters}\ }\textbf {\bibinfo {volume}
  {103}},\ \bibinfo {pages} {210401} (\bibinfo {year} {2009})}\BibitemShut
  {NoStop}%
\bibitem [{\citenamefont {Budini}(2018)}]{budini2018maximally}%
  \BibitemOpen
  \bibfield  {author} {\bibinfo {author} {\bibfnamefont {A.~A.}\ \bibnamefont
  {Budini}},\ }\bibfield  {title} {\bibinfo {title} {Maximally non-markovian
  quantum dynamics without environment-to-system backflow of information},\
  }\href@noop {} {\bibfield  {journal} {\bibinfo  {journal} {Physical Review
  A}\ }\textbf {\bibinfo {volume} {97}},\ \bibinfo {pages} {052133} (\bibinfo
  {year} {2018})}\BibitemShut {NoStop}%
\bibitem [{\citenamefont {Budini}(2022)}]{budini2022quantum}%
  \BibitemOpen
  \bibfield  {author} {\bibinfo {author} {\bibfnamefont {A.~A.}\ \bibnamefont
  {Budini}},\ }\bibfield  {title} {\bibinfo {title} {Quantum non-markovian
  environment-to-system backflows of information: Nonoperational vs.
  operational approaches},\ }\href@noop {} {\bibfield  {journal} {\bibinfo
  {journal} {Entropy}\ }\textbf {\bibinfo {volume} {24}},\ \bibinfo {pages}
  {649} (\bibinfo {year} {2022})}\BibitemShut {NoStop}%
\bibitem [{\citenamefont {Breuer}\ \emph {et~al.}(2016)\citenamefont {Breuer},
  \citenamefont {Laine}, \citenamefont {Piilo},\ and\ \citenamefont
  {Vacchini}}]{breuer2016colloquium}%
  \BibitemOpen
  \bibfield  {author} {\bibinfo {author} {\bibfnamefont {H.-P.}\ \bibnamefont
  {Breuer}}, \bibinfo {author} {\bibfnamefont {E.-M.}\ \bibnamefont {Laine}},
  \bibinfo {author} {\bibfnamefont {J.}~\bibnamefont {Piilo}},\ and\ \bibinfo
  {author} {\bibfnamefont {B.}~\bibnamefont {Vacchini}},\ }\bibfield  {title}
  {\bibinfo {title} {Colloquium: Non-{M}arkovian dynamics in open quantum
  systems},\ }\href {https://doi.org/10.1103/RevModPhys.88.021002} {\bibfield
  {journal} {\bibinfo  {journal} {Reviews of Modern Physics}\ }\textbf
  {\bibinfo {volume} {88}},\ \bibinfo {pages} {021002} (\bibinfo {year}
  {2016})}\BibitemShut {NoStop}%
\bibitem [{\citenamefont {Chin}\ \emph {et~al.}(2012)\citenamefont {Chin},
  \citenamefont {Huelga},\ and\ \citenamefont {Plenio}}]{chin2012quantum}%
  \BibitemOpen
  \bibfield  {author} {\bibinfo {author} {\bibfnamefont {A.~W.}\ \bibnamefont
  {Chin}}, \bibinfo {author} {\bibfnamefont {S.~F.}\ \bibnamefont {Huelga}},\
  and\ \bibinfo {author} {\bibfnamefont {M.~B.}\ \bibnamefont {Plenio}},\
  }\bibfield  {title} {\bibinfo {title} {Quantum metrology in non-markovian
  environments},\ }\href {https://doi.org/10.1103/PhysRevLett.109.233601}
  {\bibfield  {journal} {\bibinfo  {journal} {Physical Review Letters}\
  }\textbf {\bibinfo {volume} {109}},\ \bibinfo {pages} {233601} (\bibinfo
  {year} {2012})}\BibitemShut {NoStop}%
\bibitem [{\citenamefont {Lu}\ \emph {et~al.}(2010)\citenamefont {Lu},
  \citenamefont {Wang},\ and\ \citenamefont {Sun}}]{lu2010quantum}%
  \BibitemOpen
  \bibfield  {author} {\bibinfo {author} {\bibfnamefont {X.-M.}\ \bibnamefont
  {Lu}}, \bibinfo {author} {\bibfnamefont {X.}~\bibnamefont {Wang}},\ and\
  \bibinfo {author} {\bibfnamefont {C.~P.}\ \bibnamefont {Sun}},\ }\bibfield
  {title} {\bibinfo {title} {Quantum fisher information flow and
  non-{M}arkovian processes of open systems},\ }\href
  {https://doi.org/10.1103/PhysRevA.82.042103} {\bibfield  {journal} {\bibinfo
  {journal} {Physical Review A}\ }\textbf {\bibinfo {volume} {82}},\ \bibinfo
  {pages} {042103} (\bibinfo {year} {2010})}\BibitemShut {NoStop}%
\bibitem [{\citenamefont {Albarelli}\ \emph {et~al.}(2018)\citenamefont
  {Albarelli}, \citenamefont {Rossi}, \citenamefont {Tamascelli},\ and\
  \citenamefont {Genoni}}]{albarelli2018restoring}%
  \BibitemOpen
  \bibfield  {author} {\bibinfo {author} {\bibfnamefont {F.}~\bibnamefont
  {Albarelli}}, \bibinfo {author} {\bibfnamefont {M.~A.~C.}\ \bibnamefont
  {Rossi}}, \bibinfo {author} {\bibfnamefont {D.}~\bibnamefont {Tamascelli}},\
  and\ \bibinfo {author} {\bibfnamefont {M.~G.}\ \bibnamefont {Genoni}},\
  }\bibfield  {title} {\bibinfo {title} {Restoring {H}eisenberg scaling in
  noisy quantum metrology by monitoring the environment},\ }\href
  {https://doi.org/10.22331/q-2018-12-03-110} {\bibfield  {journal} {\bibinfo
  {journal} {Quantum}\ }\textbf {\bibinfo {volume} {2}},\ \bibinfo {pages}
  {110} (\bibinfo {year} {2018})}\BibitemShut {NoStop}%
\bibitem [{\citenamefont {Haroche}\ and\ \citenamefont
  {Kleppner}(1989)}]{haroche1989cavity}%
  \BibitemOpen
  \bibfield  {author} {\bibinfo {author} {\bibfnamefont {S.}~\bibnamefont
  {Haroche}}\ and\ \bibinfo {author} {\bibfnamefont {D.}~\bibnamefont
  {Kleppner}},\ }\bibfield  {title} {\bibinfo {title} {Cavity quantum
  electrodynamics},\ }\href@noop {} {\bibfield  {journal} {\bibinfo  {journal}
  {Physics Today}\ }\textbf {\bibinfo {volume} {42}},\ \bibinfo {pages} {24}
  (\bibinfo {year} {1989})}\BibitemShut {NoStop}%
\bibitem [{\citenamefont {Blais}\ \emph {et~al.}(2004)\citenamefont {Blais},
  \citenamefont {Huang}, \citenamefont {Wallraff}, \citenamefont {Girvin},\
  and\ \citenamefont {Schoelkopf}}]{blais2004cavity}%
  \BibitemOpen
  \bibfield  {author} {\bibinfo {author} {\bibfnamefont {A.}~\bibnamefont
  {Blais}}, \bibinfo {author} {\bibfnamefont {R.-S.}\ \bibnamefont {Huang}},
  \bibinfo {author} {\bibfnamefont {A.}~\bibnamefont {Wallraff}}, \bibinfo
  {author} {\bibfnamefont {S.~M.}\ \bibnamefont {Girvin}},\ and\ \bibinfo
  {author} {\bibfnamefont {R.~J.}\ \bibnamefont {Schoelkopf}},\ }\bibfield
  {title} {\bibinfo {title} {Cavity quantum electrodynamics for superconducting
  electrical circuits: An architecture for quantum computation},\ }\href@noop
  {} {\bibfield  {journal} {\bibinfo  {journal} {Physical Review A—Atomic,
  Molecular, and Optical Physics}\ }\textbf {\bibinfo {volume} {69}},\ \bibinfo
  {pages} {062320} (\bibinfo {year} {2004})}\BibitemShut {NoStop}%
\bibitem [{\citenamefont {Wallraff}\ \emph {et~al.}(2004)\citenamefont
  {Wallraff}, \citenamefont {Schuster}, \citenamefont {Blais}, \citenamefont
  {Frunzio}, \citenamefont {Huang}, \citenamefont {Majer}, \citenamefont
  {Kumar}, \citenamefont {Girvin},\ and\ \citenamefont
  {Schoelkopf}}]{wallraff2004strong}%
  \BibitemOpen
  \bibfield  {author} {\bibinfo {author} {\bibfnamefont {A.}~\bibnamefont
  {Wallraff}}, \bibinfo {author} {\bibfnamefont {D.~I.}\ \bibnamefont
  {Schuster}}, \bibinfo {author} {\bibfnamefont {A.}~\bibnamefont {Blais}},
  \bibinfo {author} {\bibfnamefont {L.}~\bibnamefont {Frunzio}}, \bibinfo
  {author} {\bibfnamefont {R.-S.}\ \bibnamefont {Huang}}, \bibinfo {author}
  {\bibfnamefont {J.}~\bibnamefont {Majer}}, \bibinfo {author} {\bibfnamefont
  {S.}~\bibnamefont {Kumar}}, \bibinfo {author} {\bibfnamefont {S.~M.}\
  \bibnamefont {Girvin}},\ and\ \bibinfo {author} {\bibfnamefont {R.~J.}\
  \bibnamefont {Schoelkopf}},\ }\bibfield  {title} {\bibinfo {title} {Strong
  coupling of a single photon to a superconducting qubit using circuit quantum
  electrodynamics},\ }\href@noop {} {\bibfield  {journal} {\bibinfo  {journal}
  {Nature}\ }\textbf {\bibinfo {volume} {431}},\ \bibinfo {pages} {162}
  (\bibinfo {year} {2004})}\BibitemShut {NoStop}%
\bibitem [{\citenamefont {Kono}\ \emph {et~al.}(2018)\citenamefont {Kono},
  \citenamefont {Koshino}, \citenamefont {Tabuchi}, \citenamefont {Noguchi},\
  and\ \citenamefont {Nakamura}}]{kono2018quantum}%
  \BibitemOpen
  \bibfield  {author} {\bibinfo {author} {\bibfnamefont {S.}~\bibnamefont
  {Kono}}, \bibinfo {author} {\bibfnamefont {K.}~\bibnamefont {Koshino}},
  \bibinfo {author} {\bibfnamefont {Y.}~\bibnamefont {Tabuchi}}, \bibinfo
  {author} {\bibfnamefont {A.}~\bibnamefont {Noguchi}},\ and\ \bibinfo {author}
  {\bibfnamefont {Y.}~\bibnamefont {Nakamura}},\ }\bibfield  {title} {\bibinfo
  {title} {Quantum non-demolition detection of an itinerant microwave photon},\
  }\href@noop {} {\bibfield  {journal} {\bibinfo  {journal} {Nature Physics}\
  }\textbf {\bibinfo {volume} {14}},\ \bibinfo {pages} {546} (\bibinfo {year}
  {2018})}\BibitemShut {NoStop}%
\bibitem [{\citenamefont {Gard}\ \emph {et~al.}(2018)\citenamefont {Gard},
  \citenamefont {Jacobs}, \citenamefont {Aumentado},\ and\ \citenamefont
  {Simmonds}}]{gard2018fast}%
  \BibitemOpen
  \bibfield  {author} {\bibinfo {author} {\bibfnamefont {B.~T.}\ \bibnamefont
  {Gard}}, \bibinfo {author} {\bibfnamefont {K.}~\bibnamefont {Jacobs}},
  \bibinfo {author} {\bibfnamefont {J.}~\bibnamefont {Aumentado}},\ and\
  \bibinfo {author} {\bibfnamefont {R.~W.}\ \bibnamefont {Simmonds}},\
  }\bibfield  {title} {\bibinfo {title} {Fast, high-fidelity, quantum
  non-demolition readout of a superconducting qubit using a transverse
  coupling},\ }\href@noop {} {\bibfield  {journal} {\bibinfo  {journal} {arXiv
  preprint arXiv:1809.02597}\ } (\bibinfo {year} {2018})}\BibitemShut {NoStop}%
\bibitem [{\citenamefont {Majer}\ \emph {et~al.}(2007)\citenamefont {Majer},
  \citenamefont {Chow}, \citenamefont {Gambetta}, \citenamefont {Koch},
  \citenamefont {Johnson}, \citenamefont {Schreier}, \citenamefont {Frunzio},
  \citenamefont {Schuster}, \citenamefont {Houck}, \citenamefont {Wallraff}
  \emph {et~al.}}]{majer2007coupling}%
  \BibitemOpen
  \bibfield  {author} {\bibinfo {author} {\bibfnamefont {J.}~\bibnamefont
  {Majer}}, \bibinfo {author} {\bibfnamefont {J.}~\bibnamefont {Chow}},
  \bibinfo {author} {\bibfnamefont {J.}~\bibnamefont {Gambetta}}, \bibinfo
  {author} {\bibfnamefont {J.}~\bibnamefont {Koch}}, \bibinfo {author}
  {\bibfnamefont {B.}~\bibnamefont {Johnson}}, \bibinfo {author} {\bibfnamefont
  {J.}~\bibnamefont {Schreier}}, \bibinfo {author} {\bibfnamefont
  {L.}~\bibnamefont {Frunzio}}, \bibinfo {author} {\bibfnamefont
  {D.}~\bibnamefont {Schuster}}, \bibinfo {author} {\bibfnamefont {A.~A.}\
  \bibnamefont {Houck}}, \bibinfo {author} {\bibfnamefont {A.}~\bibnamefont
  {Wallraff}}, \emph {et~al.},\ }\bibfield  {title} {\bibinfo {title} {Coupling
  superconducting qubits via a cavity bus},\ }\href@noop {} {\bibfield
  {journal} {\bibinfo  {journal} {Nature}\ }\textbf {\bibinfo {volume} {449}},\
  \bibinfo {pages} {443} (\bibinfo {year} {2007})}\BibitemShut {NoStop}%
\bibitem [{\citenamefont {Grimsmo}\ and\ \citenamefont
  {Parkins}(2013)}]{grimsmo2013cavity}%
  \BibitemOpen
  \bibfield  {author} {\bibinfo {author} {\bibfnamefont {A.~L.}\ \bibnamefont
  {Grimsmo}}\ and\ \bibinfo {author} {\bibfnamefont {S.}~\bibnamefont
  {Parkins}},\ }\bibfield  {title} {\bibinfo {title} {Cavity-qed simulation of
  qubit-oscillator dynamics in the ultrastrong-coupling regime},\ }\href@noop
  {} {\bibfield  {journal} {\bibinfo  {journal} {Physical Review A—Atomic,
  Molecular, and Optical Physics}\ }\textbf {\bibinfo {volume} {87}},\ \bibinfo
  {pages} {033814} (\bibinfo {year} {2013})}\BibitemShut {NoStop}%
\bibitem [{\citenamefont {Eickbusch}\ \emph {et~al.}(2022)\citenamefont
  {Eickbusch}, \citenamefont {Sivak}, \citenamefont {Ding}, \citenamefont
  {Elder}, \citenamefont {Jha}, \citenamefont {Venkatraman}, \citenamefont
  {Royer}, \citenamefont {Girvin}, \citenamefont {Schoelkopf},\ and\
  \citenamefont {Devoret}}]{eickbusch2022fast}%
  \BibitemOpen
  \bibfield  {author} {\bibinfo {author} {\bibfnamefont {A.}~\bibnamefont
  {Eickbusch}}, \bibinfo {author} {\bibfnamefont {V.}~\bibnamefont {Sivak}},
  \bibinfo {author} {\bibfnamefont {A.~Z.}\ \bibnamefont {Ding}}, \bibinfo
  {author} {\bibfnamefont {S.~S.}\ \bibnamefont {Elder}}, \bibinfo {author}
  {\bibfnamefont {S.~R.}\ \bibnamefont {Jha}}, \bibinfo {author} {\bibfnamefont
  {J.}~\bibnamefont {Venkatraman}}, \bibinfo {author} {\bibfnamefont
  {B.}~\bibnamefont {Royer}}, \bibinfo {author} {\bibfnamefont {S.~M.}\
  \bibnamefont {Girvin}}, \bibinfo {author} {\bibfnamefont {R.~J.}\
  \bibnamefont {Schoelkopf}},\ and\ \bibinfo {author} {\bibfnamefont {M.~H.}\
  \bibnamefont {Devoret}},\ }\bibfield  {title} {\bibinfo {title} {Fast
  universal control of an oscillator with weak dispersive coupling to a
  qubit},\ }\href@noop {} {\bibfield  {journal} {\bibinfo  {journal} {Nature
  Physics}\ }\textbf {\bibinfo {volume} {18}},\ \bibinfo {pages} {1464}
  (\bibinfo {year} {2022})}\BibitemShut {NoStop}%
\bibitem [{\citenamefont {Gisin}\ and\ \citenamefont
  {Percival}(1992)}]{gisin1992quantum}%
  \BibitemOpen
  \bibfield  {author} {\bibinfo {author} {\bibfnamefont {N.}~\bibnamefont
  {Gisin}}\ and\ \bibinfo {author} {\bibfnamefont {I.~C.}\ \bibnamefont
  {Percival}},\ }\bibfield  {title} {\bibinfo {title} {The quantum-state
  diffusion model applied to open systems},\ }\href@noop {} {\bibfield
  {journal} {\bibinfo  {journal} {Journal of Physics A: Mathematical and
  General}\ }\textbf {\bibinfo {volume} {25}},\ \bibinfo {pages} {5677}
  (\bibinfo {year} {1992})}\BibitemShut {NoStop}%
\bibitem [{\citenamefont {Percival}(1998)}]{percival1998quantum}%
  \BibitemOpen
  \bibfield  {author} {\bibinfo {author} {\bibfnamefont {I.}~\bibnamefont
  {Percival}},\ }\href@noop {} {\emph {\bibinfo {title} {Quantum state
  diffusion}}}\ (\bibinfo  {publisher} {Cambridge University Press},\ \bibinfo
  {year} {1998})\BibitemShut {NoStop}%
\bibitem [{\citenamefont {Di{\'o}si}\ \emph {et~al.}(1998)\citenamefont
  {Di{\'o}si}, \citenamefont {Gisin},\ and\ \citenamefont
  {Strunz}}]{diosi1998non}%
  \BibitemOpen
  \bibfield  {author} {\bibinfo {author} {\bibfnamefont {L.}~\bibnamefont
  {Di{\'o}si}}, \bibinfo {author} {\bibfnamefont {N.}~\bibnamefont {Gisin}},\
  and\ \bibinfo {author} {\bibfnamefont {W.~T.}\ \bibnamefont {Strunz}},\
  }\bibfield  {title} {\bibinfo {title} {Non-markovian quantum state
  diffusion},\ }\href@noop {} {\bibfield  {journal} {\bibinfo  {journal}
  {Physical Review A}\ }\textbf {\bibinfo {volume} {58}},\ \bibinfo {pages}
  {1699} (\bibinfo {year} {1998})}\BibitemShut {NoStop}%
\bibitem [{\citenamefont {Strunz}\ \emph {et~al.}(1999)\citenamefont {Strunz},
  \citenamefont {Di{\'o}si},\ and\ \citenamefont {Gisin}}]{strunz1999open}%
  \BibitemOpen
  \bibfield  {author} {\bibinfo {author} {\bibfnamefont {W.~T.}\ \bibnamefont
  {Strunz}}, \bibinfo {author} {\bibfnamefont {L.}~\bibnamefont {Di{\'o}si}},\
  and\ \bibinfo {author} {\bibfnamefont {N.}~\bibnamefont {Gisin}},\ }\bibfield
   {title} {\bibinfo {title} {Open system dynamics with non-markovian quantum
  trajectories},\ }\href {https://doi.org/10.1103/PhysRevLett.82.1801}
  {\bibfield  {journal} {\bibinfo  {journal} {Physical Review Letters}\
  }\textbf {\bibinfo {volume} {82}},\ \bibinfo {pages} {1801} (\bibinfo {year}
  {1999})}\BibitemShut {NoStop}%
\bibitem [{\citenamefont {Yu}\ \emph {et~al.}(1999)\citenamefont {Yu},
  \citenamefont {Di{\'o}si}, \citenamefont {Gisin},\ and\ \citenamefont
  {Strunz}}]{yu1999non}%
  \BibitemOpen
  \bibfield  {author} {\bibinfo {author} {\bibfnamefont {T.}~\bibnamefont
  {Yu}}, \bibinfo {author} {\bibfnamefont {L.}~\bibnamefont {Di{\'o}si}},
  \bibinfo {author} {\bibfnamefont {N.}~\bibnamefont {Gisin}},\ and\ \bibinfo
  {author} {\bibfnamefont {W.~T.}\ \bibnamefont {Strunz}},\ }\bibfield  {title}
  {\bibinfo {title} {Non-markovian quantum-state diffusion: Perturbation
  approach},\ }\href@noop {} {\bibfield  {journal} {\bibinfo  {journal}
  {Physical Review A}\ }\textbf {\bibinfo {volume} {60}},\ \bibinfo {pages}
  {91} (\bibinfo {year} {1999})}\BibitemShut {NoStop}%
\bibitem [{\citenamefont {Jaynes}\ and\ \citenamefont
  {Cummings}(1963)}]{jaynes1963comparison}%
  \BibitemOpen
  \bibfield  {author} {\bibinfo {author} {\bibfnamefont {E.~T.}\ \bibnamefont
  {Jaynes}}\ and\ \bibinfo {author} {\bibfnamefont {F.~W.}\ \bibnamefont
  {Cummings}},\ }\bibfield  {title} {\bibinfo {title} {Comparison of quantum
  and semiclassical radiation theories with application to the beam maser},\
  }\href@noop {} {\bibfield  {journal} {\bibinfo  {journal} {Proceedings of the
  IEEE}\ }\textbf {\bibinfo {volume} {51}},\ \bibinfo {pages} {89} (\bibinfo
  {year} {1963})}\BibitemShut {NoStop}%
\bibitem [{\citenamefont {Haroche}\ and\ \citenamefont
  {Raimond}(2006)}]{haroche2006exploring}%
  \BibitemOpen
  \bibfield  {author} {\bibinfo {author} {\bibfnamefont {S.}~\bibnamefont
  {Haroche}}\ and\ \bibinfo {author} {\bibfnamefont {J.-M.}\ \bibnamefont
  {Raimond}},\ }\href@noop {} {\emph {\bibinfo {title} {Exploring the Quantum:
  Atoms, Cavities, and Photons}}}\ (\bibinfo  {publisher} {Oxford University
  Press},\ \bibinfo {year} {2006})\BibitemShut {NoStop}%
\bibitem [{\citenamefont {Blais}\ \emph {et~al.}(2021)\citenamefont {Blais},
  \citenamefont {Grimsmo}, \citenamefont {Girvin},\ and\ \citenamefont
  {Wallraff}}]{blais2021circuit}%
  \BibitemOpen
  \bibfield  {author} {\bibinfo {author} {\bibfnamefont {A.}~\bibnamefont
  {Blais}}, \bibinfo {author} {\bibfnamefont {A.~L.}\ \bibnamefont {Grimsmo}},
  \bibinfo {author} {\bibfnamefont {S.~M.}\ \bibnamefont {Girvin}},\ and\
  \bibinfo {author} {\bibfnamefont {A.}~\bibnamefont {Wallraff}},\ }\bibfield
  {title} {\bibinfo {title} {Circuit quantum electrodynamics},\ }\href
  {https://doi.org/10.1103/RevModPhys.93.025005} {\bibfield  {journal}
  {\bibinfo  {journal} {Reviews of Modern Physics}\ }\textbf {\bibinfo {volume}
  {93}},\ \bibinfo {pages} {025005} (\bibinfo {year} {2021})}\BibitemShut
  {NoStop}%
\bibitem [{\citenamefont {Schuster}\ \emph {et~al.}(2007)\citenamefont
  {Schuster}, \citenamefont {Houck}, \citenamefont {Schreier}, \citenamefont
  {Wallraff}, \citenamefont {Gambetta}, \citenamefont {Blais}, \citenamefont
  {Frunzio}, \citenamefont {Majer}, \citenamefont {Johnson}, \citenamefont
  {Devoret}, \citenamefont {Girvin},\ and\ \citenamefont
  {Schoelkopf}}]{schuster2007resolving}%
  \BibitemOpen
  \bibfield  {author} {\bibinfo {author} {\bibfnamefont {D.~I.}\ \bibnamefont
  {Schuster}}, \bibinfo {author} {\bibfnamefont {A.~A.}\ \bibnamefont {Houck}},
  \bibinfo {author} {\bibfnamefont {J.~A.}\ \bibnamefont {Schreier}}, \bibinfo
  {author} {\bibfnamefont {A.}~\bibnamefont {Wallraff}}, \bibinfo {author}
  {\bibfnamefont {J.~M.}\ \bibnamefont {Gambetta}}, \bibinfo {author}
  {\bibfnamefont {A.}~\bibnamefont {Blais}}, \bibinfo {author} {\bibfnamefont
  {L.}~\bibnamefont {Frunzio}}, \bibinfo {author} {\bibfnamefont
  {J.}~\bibnamefont {Majer}}, \bibinfo {author} {\bibfnamefont
  {B.}~\bibnamefont {Johnson}}, \bibinfo {author} {\bibfnamefont {M.~H.}\
  \bibnamefont {Devoret}}, \bibinfo {author} {\bibfnamefont {S.~M.}\
  \bibnamefont {Girvin}},\ and\ \bibinfo {author} {\bibfnamefont {R.~J.}\
  \bibnamefont {Schoelkopf}},\ }\bibfield  {title} {\bibinfo {title} {Resolving
  photon number states in a superconducting circuit},\ }\href
  {https://doi.org/10.1038/nature05461} {\bibfield  {journal} {\bibinfo
  {journal} {Nature}\ }\textbf {\bibinfo {volume} {445}},\ \bibinfo {pages}
  {515} (\bibinfo {year} {2007})}\BibitemShut {NoStop}%
\bibitem [{\citenamefont {T{\'o}th}\ and\ \citenamefont
  {Petz}(2013)}]{TothPetz2013}%
  \BibitemOpen
  \bibfield  {author} {\bibinfo {author} {\bibfnamefont {G.}~\bibnamefont
  {T{\'o}th}}\ and\ \bibinfo {author} {\bibfnamefont {D.}~\bibnamefont
  {Petz}},\ }\bibfield  {title} {\bibinfo {title} {Extremal properties of the
  variance and the quantum fisher information},\ }\href
  {https://doi.org/10.1103/PhysRevA.87.032324} {\bibfield  {journal} {\bibinfo
  {journal} {Phys. Rev. A}\ }\textbf {\bibinfo {volume} {87}},\ \bibinfo
  {pages} {032324} (\bibinfo {year} {2013})}\BibitemShut {NoStop}%
\bibitem [{\citenamefont {T{\'o}th}\ and\ \citenamefont
  {Fr{\"o}wis}(2022)}]{TothFrowis2022}%
  \BibitemOpen
  \bibfield  {author} {\bibinfo {author} {\bibfnamefont {G.}~\bibnamefont
  {T{\'o}th}}\ and\ \bibinfo {author} {\bibfnamefont {F.}~\bibnamefont
  {Fr{\"o}wis}},\ }\bibfield  {title} {\bibinfo {title} {Uncertainty relations
  with the variance and the quantum fisher information based on convex
  decompositions of density matrices},\ }\href
  {https://doi.org/10.1103/PhysRevResearch.4.013075} {\bibfield  {journal}
  {\bibinfo  {journal} {Phys. Rev. Research}\ }\textbf {\bibinfo {volume}
  {4}},\ \bibinfo {pages} {013075} (\bibinfo {year} {2022})}\BibitemShut
  {NoStop}%
\bibitem [{\citenamefont {Corn}\ \emph {et~al.}(2016)\citenamefont {Corn},
  \citenamefont {Jing},\ and\ \citenamefont {Yu}}]{CornJingYu2016}%
  \BibitemOpen
  \bibfield  {author} {\bibinfo {author} {\bibfnamefont {B.}~\bibnamefont
  {Corn}}, \bibinfo {author} {\bibfnamefont {J.}~\bibnamefont {Jing}},\ and\
  \bibinfo {author} {\bibfnamefont {T.}~\bibnamefont {Yu}},\ }\bibfield
  {title} {\bibinfo {title} {Non-markovian quantum trajectory unravellings of
  entanglement},\ }\href {https://doi.org/10.26421/QIC16.5-6-5} {\bibfield
  {journal} {\bibinfo  {journal} {Quantum Information and Computation}\
  }\textbf {\bibinfo {volume} {16}},\ \bibinfo {pages} {483} (\bibinfo {year}
  {2016})}\BibitemShut {NoStop}%
\bibitem [{\citenamefont {Pezz{\`e}}\ \emph {et~al.}(2018)\citenamefont
  {Pezz{\`e}}, \citenamefont {Smerzi}, \citenamefont {Oberthaler},
  \citenamefont {Schmied},\ and\ \citenamefont {Treutlein}}]{pezze2018quantum}%
  \BibitemOpen
  \bibfield  {author} {\bibinfo {author} {\bibfnamefont {L.}~\bibnamefont
  {Pezz{\`e}}}, \bibinfo {author} {\bibfnamefont {A.}~\bibnamefont {Smerzi}},
  \bibinfo {author} {\bibfnamefont {M.~K.}\ \bibnamefont {Oberthaler}},
  \bibinfo {author} {\bibfnamefont {R.}~\bibnamefont {Schmied}},\ and\ \bibinfo
  {author} {\bibfnamefont {P.}~\bibnamefont {Treutlein}},\ }\bibfield  {title}
  {\bibinfo {title} {Quantum metrology with nonclassical states of atomic
  ensembles},\ }\href {https://doi.org/10.1103/RevModPhys.90.035005} {\bibfield
   {journal} {\bibinfo  {journal} {Reviews of Modern Physics}\ }\textbf
  {\bibinfo {volume} {90}},\ \bibinfo {pages} {035005} (\bibinfo {year}
  {2018})}\BibitemShut {NoStop}%
\bibitem [{\citenamefont {Zhong}\ \emph {et~al.}(2014)\citenamefont {Zhong},
  \citenamefont {Lu}, \citenamefont {Jing},\ and\ \citenamefont
  {Wang}}]{zhong2014optimal}%
  \BibitemOpen
  \bibfield  {author} {\bibinfo {author} {\bibfnamefont {W.}~\bibnamefont
  {Zhong}}, \bibinfo {author} {\bibfnamefont {X.-M.}\ \bibnamefont {Lu}},
  \bibinfo {author} {\bibfnamefont {X.-X.}\ \bibnamefont {Jing}},\ and\
  \bibinfo {author} {\bibfnamefont {X.}~\bibnamefont {Wang}},\ }\bibfield
  {title} {\bibinfo {title} {Optimal condition for measurement observable via
  error-propagation},\ }\href {https://doi.org/10.1088/1751-8113/47/38/385304}
  {\bibfield  {journal} {\bibinfo  {journal} {Journal of Physics A:
  Mathematical and Theoretical}\ }\textbf {\bibinfo {volume} {47}},\ \bibinfo
  {pages} {385304} (\bibinfo {year} {2014})}\BibitemShut {NoStop}%
\bibitem [{\citenamefont {Fr{\"o}wis}\ \emph {et~al.}(2015)\citenamefont
  {Fr{\"o}wis}, \citenamefont {Schmied},\ and\ \citenamefont
  {Gisin}}]{frowis2015tighter}%
  \BibitemOpen
  \bibfield  {author} {\bibinfo {author} {\bibfnamefont {F.}~\bibnamefont
  {Fr{\"o}wis}}, \bibinfo {author} {\bibfnamefont {R.}~\bibnamefont
  {Schmied}},\ and\ \bibinfo {author} {\bibfnamefont {N.}~\bibnamefont
  {Gisin}},\ }\bibfield  {title} {\bibinfo {title} {Tighter quantum uncertainty
  relations following from a general probabilistic bound},\ }\href
  {https://doi.org/10.1103/PhysRevA.92.012102} {\bibfield  {journal} {\bibinfo
  {journal} {Physical Review A}\ }\textbf {\bibinfo {volume} {92}},\ \bibinfo
  {pages} {012102} (\bibinfo {year} {2015})}\BibitemShut {NoStop}%
\bibitem [{\citenamefont {Okamoto}\ \emph {et~al.}(2012)\citenamefont
  {Okamoto}, \citenamefont {Iefuji}, \citenamefont {Oyama}, \citenamefont
  {Yamagata}, \citenamefont {Imai}, \citenamefont {Fujiwara},\ and\
  \citenamefont {Takeuchi}}]{okamoto2012experimental}%
  \BibitemOpen
  \bibfield  {author} {\bibinfo {author} {\bibfnamefont {R.}~\bibnamefont
  {Okamoto}}, \bibinfo {author} {\bibfnamefont {M.}~\bibnamefont {Iefuji}},
  \bibinfo {author} {\bibfnamefont {S.}~\bibnamefont {Oyama}}, \bibinfo
  {author} {\bibfnamefont {K.}~\bibnamefont {Yamagata}}, \bibinfo {author}
  {\bibfnamefont {H.}~\bibnamefont {Imai}}, \bibinfo {author} {\bibfnamefont
  {A.}~\bibnamefont {Fujiwara}},\ and\ \bibinfo {author} {\bibfnamefont
  {S.}~\bibnamefont {Takeuchi}},\ }\bibfield  {title} {\bibinfo {title}
  {Experimental demonstration of adaptive quantum state estimation},\ }\href
  {https://doi.org/10.1103/PhysRevLett.109.130404} {\bibfield  {journal}
  {\bibinfo  {journal} {Physical Review Letters}\ }\textbf {\bibinfo {volume}
  {109}},\ \bibinfo {pages} {130404} (\bibinfo {year} {2012})}\BibitemShut
  {NoStop}%
\bibitem [{\citenamefont {Liu}\ \emph {et~al.}(2020)\citenamefont {Liu},
  \citenamefont {Yuan}, \citenamefont {Lu},\ and\ \citenamefont
  {Wang}}]{liu2020quantum}%
  \BibitemOpen
  \bibfield  {author} {\bibinfo {author} {\bibfnamefont {J.}~\bibnamefont
  {Liu}}, \bibinfo {author} {\bibfnamefont {H.}~\bibnamefont {Yuan}}, \bibinfo
  {author} {\bibfnamefont {X.-M.}\ \bibnamefont {Lu}},\ and\ \bibinfo {author}
  {\bibfnamefont {X.}~\bibnamefont {Wang}},\ }\bibfield  {title} {\bibinfo
  {title} {Quantum fisher information matrix and multiparameter estimation},\
  }\href {https://doi.org/10.1088/1751-8121/ab5d4d} {\bibfield  {journal}
  {\bibinfo  {journal} {Journal of Physics A: Mathematical and Theoretical}\
  }\textbf {\bibinfo {volume} {53}},\ \bibinfo {pages} {023001} (\bibinfo
  {year} {2020})}\BibitemShut {NoStop}%
\bibitem [{\citenamefont {Ragy}\ \emph {et~al.}(2016)\citenamefont {Ragy},
  \citenamefont {Jarzyna},\ and\ \citenamefont
  {Demkowicz-Dobrza{\'n}ski}}]{ragy2016compatibility}%
  \BibitemOpen
  \bibfield  {author} {\bibinfo {author} {\bibfnamefont {S.}~\bibnamefont
  {Ragy}}, \bibinfo {author} {\bibfnamefont {M.}~\bibnamefont {Jarzyna}},\ and\
  \bibinfo {author} {\bibfnamefont {R.}~\bibnamefont
  {Demkowicz-Dobrza{\'n}ski}},\ }\bibfield  {title} {\bibinfo {title}
  {Compatibility in multiparameter quantum metrology},\ }\href
  {https://doi.org/10.1103/PhysRevA.94.052108} {\bibfield  {journal} {\bibinfo
  {journal} {Physical Review A}\ }\textbf {\bibinfo {volume} {94}},\ \bibinfo
  {pages} {052108} (\bibinfo {year} {2016})}\BibitemShut {NoStop}%
\bibitem [{\citenamefont {Albarelli}\ \emph {et~al.}(2019)\citenamefont
  {Albarelli}, \citenamefont {Friel},\ and\ \citenamefont
  {Datta}}]{albarelli2019holevo}%
  \BibitemOpen
  \bibfield  {author} {\bibinfo {author} {\bibfnamefont {F.}~\bibnamefont
  {Albarelli}}, \bibinfo {author} {\bibfnamefont {J.~F.}\ \bibnamefont
  {Friel}},\ and\ \bibinfo {author} {\bibfnamefont {A.}~\bibnamefont {Datta}},\
  }\bibfield  {title} {\bibinfo {title} {Evaluating the {H}olevo
  {C}ram{\'e}r--{R}ao bound for multiparameter quantum metrology},\ }\href
  {https://doi.org/10.1103/PhysRevLett.123.200503} {\bibfield  {journal}
  {\bibinfo  {journal} {Physical Review Letters}\ }\textbf {\bibinfo {volume}
  {123}},\ \bibinfo {pages} {200503} (\bibinfo {year} {2019})}\BibitemShut
  {NoStop}%
\bibitem [{\citenamefont {{\v{S}}afr{\'a}nek}(2018)}]{safranek2018simple}%
  \BibitemOpen
  \bibfield  {author} {\bibinfo {author} {\bibfnamefont {D.}~\bibnamefont
  {{\v{S}}afr{\'a}nek}},\ }\bibfield  {title} {\bibinfo {title} {Simple
  expression for the quantum fisher information matrix},\ }\href
  {https://doi.org/10.1103/PhysRevA.97.042322} {\bibfield  {journal} {\bibinfo
  {journal} {Physical Review A}\ }\textbf {\bibinfo {volume} {97}},\ \bibinfo
  {pages} {042322} (\bibinfo {year} {2018})}\BibitemShut {NoStop}%
\bibitem [{\citenamefont {Rath}\ \emph {et~al.}(2021)\citenamefont {Rath},
  \citenamefont {Branciard}, \citenamefont {Minguzzi},\ and\ \citenamefont
  {Vermersch}}]{rath2021randomized}%
  \BibitemOpen
  \bibfield  {author} {\bibinfo {author} {\bibfnamefont {A.}~\bibnamefont
  {Rath}}, \bibinfo {author} {\bibfnamefont {C.}~\bibnamefont {Branciard}},
  \bibinfo {author} {\bibfnamefont {A.}~\bibnamefont {Minguzzi}},\ and\
  \bibinfo {author} {\bibfnamefont {B.}~\bibnamefont {Vermersch}},\ }\bibfield
  {title} {\bibinfo {title} {Quantum fisher information from randomized
  measurements},\ }\href {https://doi.org/10.1103/PhysRevLett.127.260501}
  {\bibfield  {journal} {\bibinfo  {journal} {Physical Review Letters}\
  }\textbf {\bibinfo {volume} {127}},\ \bibinfo {pages} {260501} (\bibinfo
  {year} {2021})}\BibitemShut {NoStop}%
\bibitem [{\citenamefont {Yu}\ \emph {et~al.}(2021)\citenamefont {Yu},
  \citenamefont {Li}, \citenamefont {Wang}, \citenamefont {Chu}, \citenamefont
  {Yang}, \citenamefont {Gong}, \citenamefont {Goldman},\ and\ \citenamefont
  {Cai}}]{yu2021experimental}%
  \BibitemOpen
  \bibfield  {author} {\bibinfo {author} {\bibfnamefont {M.}~\bibnamefont
  {Yu}}, \bibinfo {author} {\bibfnamefont {D.}~\bibnamefont {Li}}, \bibinfo
  {author} {\bibfnamefont {J.}~\bibnamefont {Wang}}, \bibinfo {author}
  {\bibfnamefont {Y.}~\bibnamefont {Chu}}, \bibinfo {author} {\bibfnamefont
  {P.}~\bibnamefont {Yang}}, \bibinfo {author} {\bibfnamefont {M.}~\bibnamefont
  {Gong}}, \bibinfo {author} {\bibfnamefont {N.}~\bibnamefont {Goldman}},\ and\
  \bibinfo {author} {\bibfnamefont {J.}~\bibnamefont {Cai}},\ }\bibfield
  {title} {\bibinfo {title} {Experimental estimation of the quantum fisher
  information from randomized measurements},\ }\href
  {https://doi.org/10.1103/PhysRevResearch.3.043122} {\bibfield  {journal}
  {\bibinfo  {journal} {Physical Review Research}\ }\textbf {\bibinfo {volume}
  {3}},\ \bibinfo {pages} {043122} (\bibinfo {year} {2021})}\BibitemShut
  {NoStop}%
\bibitem [{\citenamefont {Cerezo}\ \emph {et~al.}(2021)\citenamefont {Cerezo},
  \citenamefont {Sone}, \citenamefont {Beckey},\ and\ \citenamefont
  {Coles}}]{cerezo2021sub}%
  \BibitemOpen
  \bibfield  {author} {\bibinfo {author} {\bibfnamefont {M.}~\bibnamefont
  {Cerezo}}, \bibinfo {author} {\bibfnamefont {A.}~\bibnamefont {Sone}},
  \bibinfo {author} {\bibfnamefont {J.~L.}\ \bibnamefont {Beckey}},\ and\
  \bibinfo {author} {\bibfnamefont {P.~J.}\ \bibnamefont {Coles}},\ }\bibfield
  {title} {\bibinfo {title} {Sub-quantum fisher information},\ }\href
  {https://doi.org/10.1088/2058-9565/abfbef} {\bibfield  {journal} {\bibinfo
  {journal} {Quantum Science and Technology}\ }\textbf {\bibinfo {volume}
  {6}},\ \bibinfo {pages} {035008} (\bibinfo {year} {2021})}\BibitemShut
  {NoStop}%
\end{thebibliography}
\end{document}